\documentclass[11pt]{article}

\usepackage[margin=1in]{geometry}
\usepackage[nodisplayskipstretch]{setspace}
\usepackage{booktabs}

\usepackage{amsmath,amsfonts,amssymb}
\usepackage{amsthm}
\usepackage{bm}

\usepackage{graphicx}
\usepackage{float}
\usepackage{subcaption}

\usepackage{algorithm}
\usepackage{algpseudocode}

\usepackage{mathptmx}
\DeclareMathAlphabet{\mathcal}{OMS}{cmsy}{m}{n}
\usepackage{multirow}
\usepackage{enumitem}
\usepackage{xurl}
\usepackage{xcolor}

\usepackage[numbers,sort&compress]{natbib}
\usepackage{hyperref}
\hypersetup{
	colorlinks=true,
	linkcolor=blue,
	citecolor=blue,
	urlcolor=blue
}

\DeclareMathAlphabet\mathbfcal{OMS}{cmsy}{b}{n}

\newcommand{\diag}{\text{diag}}

\newcommand{\tr}{\text{tr}}

\newcommand{\bQ}{\mathbf{Q}}

\title{multiCAsp: Spatially Informed Canonical Association Analysis of Multi-Omic Data}

\author{
	Ziqi Mu$^{1}$ \and
	Qiuyi Wu$^{2}$ \and
	Zhicheng Ji$^{3}$ \and
	Pixu Shi$^{3,4*}$
}

\date{}

\begin{document}
	\doublespacing
	
	\maketitle
	
	\begin{center}
		{\small
			$^{1}$Department of Statistics, Duke University, Durham, NC 27710, United States\\
			$^{2}$Department of Epidemiology and Biostatistics, Temple University, Philadelphia, PA 19122, United States\\
			$^{3}$Department of Biostatistics \& Bioinformatics, Duke University, Durham, NC 27710, United States\\
			$^{4}$Duke Microbiome Center, Duke University School of Medicine, Durham, NC 27710, United States\\[4pt]
			$^{*}$\textit{Correspondence:}
			\href{mailto:pixu.shi@duke.edu}{pixu.shi@duke.edu}
		}
	\end{center}
	
	
	\begin{abstract}
		Simultaneous profiling of multiple molecular layers at spatially resolved locations creates opportunities to identify shared cross-modal biological variation in its tissue context. We present multiCAsp (\textbf{multi}-omic \textbf{CA}nonical Association informed by \textbf{sp}atial graph), a probabilistic latent-factor model for integrating $M \geq 2$ spatially resolved omic modalities. It incorporates spatial dependence through a graph-Laplacian prior on shared latent factors and uses an expectation-maximization algorithm to estimate factor scores and modality-specific feature loadings. Because the prior is defined on a neighborhood graph rather than on Euclidean distances, it accommodates irregular tissue geometries and extends naturally to multiple tissue sections. Simulations show that multiCAsp consistently achieves higher mean latent subspace and loading recovery than alternative methods across varying sample sizes, feature dimensions, and noise levels, with the largest gains under high noise. Compared with an existing Gaussian process–based spatial factor model, it is also less sensitive to how the spatial coordinates are represented, produces spatially smoother factors, and has lower computational cost. Applied to metabolomic-proteomic data from macaque cortex and gene-protein data from human renal carcinoma, multiCAsp recovers spatially coherent multi-omic factors with interpretable biological signatures. In the macaque data, transferred factors retain broad spatial patterns across animals and support cross-modal prediction of protein abundances from metabolomic measurements.

	\end{abstract}
	
	
	
	\section{Background}
	\label{sec:background}
	
	Advances in spatially resolved genomics have made it possible to profile multiple molecular layers simultaneously at defined tissue locations. Imaging-based platforms such as MERFISH and 10x Xenium In Situ, and sequencing-based platforms such as 10x Visium, capture gene expression, protein abundance, and other molecular features alongside their spatial coordinates within tissue sections. A distinct paradigm, suited to spatial scales too large for imaging-based approaches, collects anatomically mapped tissue samples from defined locations across an organ and profiles each sample through multiple assay platforms, yielding multi-omic measurements with explicit spatial coordinates. One example is Wu et al.~\cite{wu2026integrated}, where metabolites and proteins were profiled from dissected cortical hemispheres of adult rhesus monkeys. This design is analogous to ecology-type spatial sampling, in which biological gradients are inferred from community measurements across spatial units, typically yielding fewer units, less precise spatial positions, and noisier measurements than imaging- or sequencing-based platforms. Across both paradigms, a central challenge is to identify cross-modal molecular programs whose variation is spatially organized, as such patterns reveal how tissue architecture shapes multi-omic phenotypes.
	
	Canonical correlation analysis (CCA) and its regularized variants, including sparse CCA \cite{witten2009penalized}, are natural tools for identifying shared variation between two high-dimensional data matrices. Multi-view factor analysis methods such as MOFA and MOFA+ \cite{argelaguet2018multi,argelaguet2020mofa} generalize CCA to more than two modalities by simultaneously decomposing multiple data matrices into shared and modality-specific latent factors. However, these approaches treat observations as independent, ignoring spatial proximity or tissue graph structure, so recovered factors may reflect covariation driven by technical effects rather than spatially organized biology. Spatial extensions of CCA, such as canonical spatial correlation analysis \cite{lin2025canonical} and a graph-Laplacian-regularized sparse CCA \cite{wu2026integrated}, address this problem, but remain restricted to two modalities.
	
	Methods that incorporate spatial structure have been developed primarily for single-modality data. SPARK \cite{sun2020statistical}, BayesSpace \cite{zhao2021spatial}, and SpatialPCA \cite{shang2022spatially} incorporate spatial covariance into differential expression analysis or dimensionality reduction for a single molecular layer. Spatially constrained PCA for ecological genetics \cite{jombart2008revealing} applies a similar idea to genetic variation data. MEFISTO extends MOFA+ with Gaussian process priors to incorporate spatial covariance \cite{velten2022identifying}, though it only demonstrated spatial application to single-modality data and the multi-omic application it presents uses temporal rather than spatial structure. In addition, MEFISTO represents spatial dependence through a kernel applied to Euclidean coordinates, which is computationally demanding in practice and is a poor proxy for tissue adjacency in anatomically complex structures where biological boundaries do not align with coordinate distance. For multi-modal data collected at irregular or three-dimensional locations, such as a dissected cortical hemisphere, no existing probabilistic factor model, to our knowledge, captures spatial covariance across modalities through tissue adjacency rather than Euclidean distance. Recovering shared cross-modal patterns that follow tissue structure, without smoothing across anatomical boundaries or depending on coordinate assignment, remains a methodological gap.
	
	Here we introduce multiCAsp, a unified probabilistic model for spatially resolved multi-omic data with $M \geq 2$ modalities. multiCAsp places a graph-Laplacian-based Gaussian prior on the shared latent factors, encouraging each component to vary smoothly over space while drawing information from all modalities simultaneously. To handle high-dimensional measurements, multiCAsp applies modality-specific PCA where needed before jointly estimating the shared spatial factors and loadings from the reduced representations. Because the spatial prior acts directly on the joint cross-modal factor matrix, the model can connect features from different molecular layers that share the same underlying spatial organization, even when individual modality signals are noisy. Unlike methods that model spatial dependence through Euclidean-distance kernels, the graph Laplacian represents spatial relationships through a neighborhood graph. This allows multiCAsp to handle three-dimensional tissues that are difficult to flatten without stretching or cutting, and irregular geometries where samples that are close in space could be physically separated. It also accommodates multiple independent spatial maps and is less sensitive to how spatial coordinates are assigned.
	
	The contributions of this work are as follows. We introduce multiCAsp, a probabilistic model that unifies canonical association analysis across $M\geq 2$ modalities, with a principled spatial Gaussian prior on the shared latent factors to reflect spatial adjacency of the samples. We develop an efficient EM algorithm that exploits the Kronecker structure of the posterior, with closed-form updates for the loading matrices and noise variances, and the spatial smoothing parameter estimated jointly. In simulation studies, we show that multiCAsp improves subspace and loading recovery and produces spatially smoother factors than competing methods. Compared with MEFISTO, a widely used alternative based on Euclidean distance, multiCAsp is also more stable under nonlinear transformation of the coordinate system and runs substantially faster. Finally, we demonstrate the utility of multiCAsp on two spatially resolved multi-omic datasets: spatially mapped metabolomic and proteomic profiles of dissected macaque cortex \cite{wu2026integrated} and co-profiled gene expression and protein abundance from human renal cell carcinoma tissue \cite{10xgenomics2025xenium}, illustrating how the method identifies spatially organized cross-modal molecular patterns with biologically interpretable factor structures. Using the macaque data, we also evaluate cross-animal transfer of spatial factors and cross-modal prediction of protein abundance from metabolomic data.

	
	\section{Methods}
	\label{sec:methods}
	
	\subsection{Model setting}
	\label{sec:model-setting}
	
	\paragraph{Sample spatial structure.}
	Suppose there are $n$ samples with a spatial neighborhood matrix $W_0 \in \mathbb{R}^{n \times n}$ defining their adjacency, where $(W_0)_{ij}=1$ if samples $i$ and $j$ are spatial neighbors and 0 otherwise. Such neighborhood matrices can be obtained by Delaunay triangulation on the samples’ spatial coordinates, in two or three dimensions, or through manual curation. We assume each sample has at least one neighbor. Let $D$ be the diagonal degree matrix of $W_0$, and define the normalized graph Laplacian as 
	\begin{equation}\label{eq:laplacian}
	L = I - D^{-1/2} W_0 D^{-1/2}.
	\end{equation}
	Let $L = QVQ^\top$ be the eigendecomposition of $L$, where $Q$ contains the orthonormal eigenvectors and $V=\operatorname{diag}(v_1,\ldots,v_n)$ contains the corresponding eigenvalues. We then construct a spatial smoothing matrix 
	\begin{equation}\label{eq:matrix_W}
	W = (\eta L + I)^{-1} = Q(\eta V + I)^{-1} Q^\top,
	\end{equation}
	where $\eta \geq 0$ is a smoothing parameter that controls the amount of spatial smoothing. Because $W$ is defined through the neighborhood graph rather than Euclidean distances, it accommodates complex tissue geometries, such as folded or three-dimensional tissues that are difficult to flatten and samples from multiple tissue sections. This framework also accommodates other positive definite constructions of $W$, such as a Gaussian kernel on Euclidean distances, although the update for $\eta$ in Section~\ref{sec:model-estimation} is specific to the graph-Laplacian construction used throughout this paper.
	
	\paragraph{Multi-modal latent representation.}
	Let $M \geq 2$ denote the number of data modalities, and let $X^{(m)} \in \mathbb{R}^{n \times p_m}$ denote the data matrix of modality $m$, with $p_m$ features measured on the same $n$ spatially resolved samples. multiCAsp models each modality as
	\begin{equation}\label{eq:X_model}
	X^{(m)} = S A^{(m)} P^{(m)\top} + \sigma_m E^{(m)}, \qquad E^{(m)} \sim \mathcal{N}_{n,p_m}(0, I_n, I_{p_m}), \qquad m = 1, \ldots, M.
	\end{equation}
	Here $S = [s_1, \ldots, s_r] \in \mathbb{R}^{n \times r}$ contains $r$ shared spatial latent factors, common to all modalities, whose values at the $n$ samples describe spatial patterns of cross-modal variation. Each factor has an independent spatial prior
	\begin{equation}\label{eq:s_prior}
	s_k \sim \mathcal{N}_n(0, W), \qquad k = 1, \ldots, r,
	\end{equation}
	with $W$ defined in \eqref{eq:matrix_W}. Under this prior, values of each factor at neighboring samples are positively correlated for $\eta > 0$, so factors tend to vary smoothly across the spatial graph, and larger $\eta$ enforces stronger smoothing, with $\eta = 0$ corresponding to independent samples. The matrix $P^{(m)} \in \mathbb{R}^{p_m \times d_m}$, with $d_m \ll p_m$, is a modality-specific loading basis satisfying $P^{(m)\top} P^{(m)} = I_{d_m}$. It represents each modality in a few dimensions, which handles the high dimensionality and collinearity of features that commonly arise in multi-omic data. In practice, $P^{(m)}$ is obtained by modality-specific PCA, or set to the identity when $p_m$ is small, and held fixed during estimation. The matrix $A^{(m)} \in \mathbb{R}^{r \times d_m}$ is the dimension-reduced loading matrix of modality $m$, which describes how each shared factor is expressed in that modality, and $A^{(m)} P^{(m)\top} \in \mathbb{R}^{r \times p_m}$ gives the corresponding loadings on the original features, used for biological interpretation. Finally, $\sigma_m^2$ is the noise variance of modality $m$, which allows modalities with different measurement precision to contribute different amounts of information to the shared factors.
	
	Equivalently, writing $S = W^{1/2} S_0$ with $S_0 \sim \mathcal{N}_{n,r}(0, I_n, I_r)$,
	\begin{equation}\label{eq:X_full_model}
	X^{(m)} = W^{1/2} S_0 A^{(m)} P^{(m)\top} + \sigma_m E^{(m)}.
	\end{equation}

	\subsection{Model estimation}
	\label{sec:model-estimation}
	
	The parameters are estimated using an EM algorithm, treating the shared spatial factor matrix $S$ as the latent variable and \[ \theta = \bigl(\{\sigma_m^2\}_{m=1}^M,\; \eta,\; \{A^{(m)}\}_{m=1}^M\bigr) \] as parameters.
	
	\paragraph{Projected data.}
	For each modality, define the projected data matrix \[ \tilde{X}^{(m)} = X^{(m)} P^{(m)} \in \mathbb{R}^{n \times d_m}. \] Since $P^{(m)\top} P^{(m)} = I_{d_m}$, the model reduces to \[ \tilde{X}^{(m)} = S A^{(m)} + \sigma_m \tilde{E}^{(m)}, \qquad \tilde{E}^{(m)} \sim \mathcal{N}_{n,d_m}(0, I_n, I_{d_m}), \qquad m = 1, \ldots, M. \]
	
	\paragraph{Joint log-likelihood.}
	The conditional likelihoods are \[ p(\tilde{X}^{(m)} \mid S) \propto (\sigma_m^2)^{-nd_m/2} \exp\!\left\{ -\frac{1}{2\sigma_m^2} \mathrm{vec}(\tilde{X}^{(m)} - S A^{(m)})^\top \mathrm{vec}(\tilde{X}^{(m)} - S A^{(m)}) \right\}, \] and the prior on $S$ is \[ p(S) \propto |W|^{-r/2} \exp\!\left( -\frac{1}{2}\,\mathrm{vec}(S)^\top (I_r \otimes W^{-1})\,\mathrm{vec}(S) \right). \]
	
	Since the modalities are conditionally independent given $S$, the joint log-likelihood is \[ \log p\!\left(\tilde{X}^{(1)}, \ldots, \tilde{X}^{(M)}, S\right) = \log p(S) + \sum_{m=1}^{M} \log p(\tilde{X}^{(m)} \mid S). \]
	
	Reorganizing in terms of $S$, the quadratic form in $\mathrm{vec}(S)$ is determined by
	\begin{align*}
	H &= H\!\left(\{\sigma_m^2\}, \{A^{(m)}\}\right) = \sum_{m=1}^{M} \frac{1}{\sigma_m^2}\, \tilde{X}^{(m)} A^{(m)\top} \;\in \mathbb{R}^{n \times r}, \\ \bQ &= \bQ\!\left(\{\sigma_m^2\}, \{A^{(m)}\}, \eta\right) = \sum_{m=1}^{M} \frac{1}{\sigma_m^2}\, A^{(m)} A^{(m)\top} \otimes I_n + I_r \otimes W^{-1} \;\in \mathbb{R}^{nr \times nr}.
	\end{align*}
	
	Here, $H$ combines information from all modalities, weighted by their noise variances. The posterior precision matrix $\bQ$ has a Kronecker-sum structure that combines information from the modalities with the spatial precision $W^{-1}$. Writing $B=\sum_{m}\sigma_m^{-2}A^{(m)}A^{(m)\top}=U\Lambda U^\top$ and $W^{-1}=\eta L+I=Q(\eta V+I)Q^\top$, we have $\bQ=(U\otimes Q)\{\Lambda\otimes I_n+I_r\otimes(\eta V+I)\}(U\otimes Q)^\top$ with a diagonal middle factor, so $\bQ^{-1}$ requires only the eigendecomposition of $L$, computed once, and an $r\times r$ eigendecomposition per iteration.
	
	\paragraph{Posterior distribution of $S$.}
	The posterior is \[ \mathrm{vec}(S) \mid \tilde{X}^{(1)}, \ldots, \tilde{X}^{(M)} \;\sim\; \mathcal{N}\!\left(\bQ^{-1}\mathrm{vec}(H),\; \bQ^{-1}\right). \] Denote the posterior mean matrix as $\mathcal{M}^{(t)} = \mathrm{mat}\!\left(\bQ^{-1}\mathrm{vec}(H)\right) \in \mathbb{R}^{n \times r}$ and the posterior covariance as $\bm{\Sigma}^{(t)} = \bQ^{-1} \in \mathbb{R}^{nr \times nr}$. Partition $\bm{\Sigma}^{(t)}$ into $n \times n$ blocks $\Sigma^{(t)}_{k\ell}$, and define \[ T^{(t)}_{k\ell} = \tr\!\left(\Sigma^{(t)}_{k\ell}\right), \qquad k, \ell = 1, \ldots, r. \]
	
	\paragraph{E-step.}
	Given the updated noise variances $\{\sigma_m^{2(t)}\}_{m=1}^{M}$
	and the previous loading matrices $\{A^{(m)(t-1)}\}_{m=1}^{M}$
	and spatial parameter $\eta^{(t-1)}$,
	compute the posterior moments:
	\[ \mathcal{M}^{(t)} = \mathrm{mat}\!\left((\bQ^{(t)})^{-1}\mathrm{vec}(H^{(t)})\right), \qquad \bm{\Sigma}^{(t)} = (\bQ^{(t)})^{-1}, \]
	where $H^{(t)}$ and $\bQ^{(t)}$ are evaluated using
	$\{\sigma_m^{2(t)}\}_{m=1}^{M}$,
	$\{A^{(m)(t-1)}\}_{m=1}^{M}$,
	and $\eta^{(t-1)}$.
	
	\paragraph{M-step Part 1: update $\eta$ and $\{A^{(m)}\}$.}
	
	\textit{Update for $\eta$:} The terms in $\mathbb{E}_t[\log p(S \mid \eta)]$
	that depend on $\eta$ are \[ \mathcal{Q}(\eta) = \frac{r}{2}\sum_{i=1}^n \log(\eta v_i + 1) - \frac{1}{2}\,\tr\!\left((\eta L + I)\,\mathbb{E}_t[SS^\top]\right), \] where $v_1, \ldots, v_n$ are the eigenvalues of $L$ and
	\[
	\mathbb{E}_t[SS^\top]
	=
	\mathcal{M}^{(t)}\mathcal{M}^{(t)\top}
	+
	\sum_{k=1}^{r}\Sigma_{kk}^{(t)},
	\]
	where $\mathbb{E}_t$ denotes expectation under the
	posterior distribution computed in the current E-step.
	
	Taking the derivative and setting it to zero gives
	\[
	\frac{\partial\mathcal{Q}}{\partial\eta}
	=
	\frac{r}{2}
	\sum_{i=1}^{n}
	\frac{v_i}{\eta v_i+1}
	-
	\frac{1}{2}
	\tr\!\left(
	L\,\mathbb{E}_t[SS^\top]
	\right)
	=0,
	\]
	which is monotone decreasing in $\eta$ and can be solved by bisection or Newton--Raphson. If the derivative is nonpositive at $\eta=0$, we set $\eta^{(t)}=0$, corresponding to no spatial smoothing.
	
	\textit{Update for $A^{(m)}$:} For each modality $m = 1, \ldots, M$,
	\begin{equation}\label{eq:update_A}
	A^{(m)(t)}=\left(\mathbb{E}_t[S^\top S]\right)^{-1}\mathbb{E}_t[S^\top]\tilde{X}^{(m)}=\left(\mathcal{M}^{(t)\top}\mathcal{M}^{(t)}+T^{(t)}\right)^{-1}
	\mathcal{M}^{(t)\top}\tilde{X}^{(m)}.
	\end{equation}
	
	\paragraph{M-step Part 2: update $\{\sigma_m^2\}$.}
	For each modality $m = 1, \ldots, M$,
	\begin{equation}\label{eq:update_sigma}
	\begin{split}
	(\sigma_m^2)^{(t)} =\; & \frac{1}{n d_m} \Bigl[ \tr\!\left(\tilde{X}^{(m)} \tilde{X}^{(m)\top}\right) - 2\,\tr\!\left(\mathcal{M}^{(t-1)} A^{(m)(t-1)} \tilde{X}^{(m)\top}\right)\\ & + \tr\!\left(\mathcal{M}^{(t-1)} A^{(m)(t-1)} A^{(m)(t-1)\top} \mathcal{M}^{(t-1)\top}\right) + \tr\!\left(A^{(m)(t-1)} A^{(m)(t-1)\top} T^{(t-1)}\right) \Bigr].
	\end{split}
	\end{equation}
	Each modality's noise variance is estimated from the residual between $\tilde{X}^{(m)}$ and its reconstruction $\mathcal{M}^{(t-1)} A^{(m)(t-1)}$, accounting for posterior uncertainty in $S$.
	
	\paragraph{Initialization.}
	The algorithm assumes $r$ is given. For each modality \(m\), initialize \(A^{(m)(0)}\) using the top \(r\) right singular vectors of the projected data matrix \(\tilde X^{(m)}=X^{(m)}P^{(m)}\). Initialize $\eta^{(0)} = 1/\mathrm{median}(\diag(V))$. Initialize $\sigma_m^{2(0)}$ using the residual variance from the initial loading fit. At iteration $t$, the noise variances $\{\sigma_m^{2(t)}\}$
	are updated using the previous posterior moments,
	followed by the E-step and updates of
	$\eta^{(t)}$ and $\{A^{(m)(t)}\}$.
	
	The full pseudocode is given in Algorithm~\ref{alg:spatial_em_M}.
	

	\par\medskip
	
	\noindent
	\begin{minipage}{\linewidth}
		\captionsetup{
			hypcap=false,
			justification=raggedright,
			singlelinecheck=false,
			aboveskip=0pt,
			belowskip=0pt
		}
		
		\hrule height 0.8pt
		\vspace{6pt}
		
		\captionof{algorithm}{EM algorithm for multiCAsp with $M$ modalities}
		\label{alg:spatial_em_M}
		
		\par\vspace{6pt}
		\hrule height 0.45pt
	\end{minipage}
	
	\nopagebreak[4]

	\begin{algorithmic}
		
		\Require Data matrices $X^{(m)}\in\mathbb{R}^{n\times p_m}$,
		projection matrices $P^{(m)}\in\mathbb{R}^{p_m\times d_m}$
		for $m=1,\ldots,M$; rank $r$; graph Laplacian $L$
		
		\Ensure Loadings $\{\hat A^{(m)}\}_{m=1}^M$,
		spatial parameter $\hat\eta$, noise variances
		$\{\hat\sigma_m^2\}_{m=1}^M$, and shared factors $\hat S$
		
		\State Compute $\tilde X^{(m)}=X^{(m)}P^{(m)}$ for each $m$
		\State Initialize $A^{(m)(0)}$ from the top $r$ right singular vectors of $\tilde X^{(m)}$
		\State Initialize $\eta^{(0)}=1/\mathrm{median}(\diag(V))$ and $\sigma_m^{2(0)}$ from residual variances
		\State Compute initial posterior moments $\mathcal{M}^{(0)}$, $\bm\Sigma^{(0)}$, and $T^{(0)}$
		\State Set $t\gets1$
		
		\Repeat
		
		\Statex \textbf{M-step Part 2: Noise variances}
		\For{$m=1,\ldots,M$}
		\State Update $\sigma_m^{2(t)}$ via~\eqref{eq:update_sigma}
		\EndFor
		
		\Statex \textbf{E-step: Posterior moments}
		\State Form
		\[
		\begin{aligned}
		H^{(t)}
		&=\sum_{m=1}^{M}
		\frac{\tilde X^{(m)}A^{(m)(t-1)\top}}
		{\sigma_m^{2(t)}},\\
		\bQ^{(t)}
		&=\sum_{m=1}^{M}
		\frac{A^{(m)(t-1)}A^{(m)(t-1)\top}}
		{\sigma_m^{2(t)}}\otimes I_n\\
		&\quad+I_r\otimes W(\eta^{(t-1)})^{-1}.
		\end{aligned}
		\]
		\State Compute $\mathcal{M}^{(t)}=\mathrm{mat}((\bQ^{(t)})^{-1}
		\mathrm{vec}(H^{(t)}))$ and
		$\bm\Sigma^{(t)}=(\bQ^{(t)})^{-1}$
		\State Set $\hat S^{(t)}=\mathcal{M}^{(t)}$ and
		$T_{k\ell}^{(t)}=\tr(\Sigma_{k\ell}^{(t)})$
		
		\Statex \textbf{M-step Part 1: Spatial parameter and loadings}
		\State Update $\eta^{(t)}$ by solving
		\[
		\frac{r}{2}\sum_{i=1}^{n}
		\frac{v_i}{\eta v_i+1}
		-\frac{1}{2}\tr\!\left(
		L\,\mathbb E_t[SS^\top]\right)=0.
		\]
		\For{$m=1,\ldots,M$}
		\State Update $A^{(m)(t)}$ via~\eqref{eq:update_A}
		\EndFor
		
		\State Set $t\gets t+1$
		
		\Until{convergence of the observed-data log-likelihood
			or relative change in parameters}
		
		\State \Return $\{\hat A^{(m)}\}_{m=1}^M$,
		$\hat\eta$, $\{\hat\sigma_m^2\}_{m=1}^M$, $\hat S$
		
	\end{algorithmic}

	\par\nobreak\vspace{-0.35\baselineskip}
	\noindent\rule{\linewidth}{0.8pt}
	\par\medskip
	
	
	\subsection{Rotation and scaling of \texorpdfstring{$S$}{S} and
		\texorpdfstring{$\{A^{(m)}\}$}{A(m)}}
	\label{sec:rotation}
	
	Let $\hat{S}$ denote the posterior mean estimate of $S$ at convergence, and define $\hat{S}_0 = W^{-1/2} \hat{S}$.
	
	To stabilize scale and remove empirical correlations among latent components, we compute the sample covariance 
	\[ \widehat{\Sigma}_S = \frac{1}{n-1} \hat{S}_0^\top \!\left(I_n - \frac{1}{n}\mathbf{1}_n\mathbf{1}_n^\top\right) \hat{S}_0 \in \mathbb{R}^{r \times r}, \]
	and whiten the latent factors: 
	\[ \tilde{S}_0 = \hat{S}_0\, \widehat{\Sigma}_S^{-1/2}. \] 
	All loading matrices are then transformed correspondingly by
	\[ \tilde{A}^{(m)} = \widehat{\Sigma}_S^{1/2} A^{(m)}, \qquad m = 1, \ldots, M, \] 
	so that $\hat{S}_0 A^{(m)} = \tilde{S}_0 \tilde{A}^{(m)}$ for each modality.
	
	Because any orthogonal rotation of the whitened factors leaves the fit unchanged, we fix the rotation by ordering factors by their cross-modal agreement, in the spirit of canonical correlation analysis. Let $G^{(m)} = \tilde X^{(m)}\tilde A^{(m)\top}/\hat\sigma_m^2 \in \mathbb{R}^{n\times r}$ denote the column-centered modality-specific factor scores, and let $\Omega = \sum_{m<l}\bigl(G^{(m)\top}G^{(l)} + G^{(l)\top}G^{(m)}\bigr)/2$. With eigendecomposition $\Omega = R\Delta R^\top$, eigenvalues in decreasing order, we define
	\[ S_{0,\mathrm{final}} = \tilde S_0 R, \qquad S_{\mathrm{final}} = W^{1/2} S_{0,\mathrm{final}}, \qquad A^{(m)}_{\mathrm{final}} = R^\top \tilde A^{(m)}, \quad m = 1, \ldots, M. \]
	This preserves the fitted signal, keeps the factors uncorrelated with unit sample variance, and is unique up to sign when the eigenvalues are distinct. We fix the sign of each factor so that its largest-magnitude loading is positive. Each column of $S_{\mathrm{final}}$ represents a shared spatial latent component, and the corresponding rows of $\{A_{\mathrm{final}}^{(m)}\}$ describe its modality-specific feature loadings.

	\section{Results}
	\label{sec:results}
	
	\subsection{Simulation studies}
	\label{sec:sim-results}
	
	We used simulations to evaluate whether spatial regularization improves the accuracy of shared latent factor recovery and per-modality feature loading estimation, and how recovered components compare in spatial structure across methods. We generated $r=3$ shared factors from the spatial prior \eqref{eq:s_prior} on a realistic irregular graph of $n=233$ spatial locations from the macaque cortical voxel dataset analyzed in Section~\ref{sec:results-a26}. Following model \eqref{eq:X_model}, we generated two modalities through low-dimensional loadings embedded in block-structured, collinear raw feature spaces via approximately orthonormal bases to mimic groups of co-regulated features, such as genes or proteins in shared pathways, that are typical of omic data. Starting from a baseline of $p_1=300$, $p_2=600$, $\sigma=0.8$, and $n=233$, we varied the feature dimension $p_1$, the noise level $\sigma$, and the sample size $n$ one at a time, with 50 replicates per setting. Full details are given in Appendix~\ref{app:sim}.
	
	We compared multiCAsp with three alternatives. MEFISTO \cite{velten2022identifying} is the primary spatial comparator. Like multiCAsp, it is a probabilistic latent factor model applicable to $M \geq 2$ modalities, but it incorporates spatial covariance through Gaussian process priors on Euclidean coordinates, so the comparison between the two reflects their different choices in spatial priors. Spatially regularized CCA variants~\cite{lin2025canonical,wu2026integrated} optimize a spatially adjusted correlation criterion without an underlying generative model and therefore were not included. The other two comparators are non-spatial CCA-based methods, one combining CCA with a PCA step and the other inducing sparsity through $\ell_1$ penalization \cite{witten2009penalized}, since multi-omic assay data often have high-dimensional and collinear features due to shared functional pathways and co-regulation, which makes direct canonical association analysis numerically unstable. MEFISTO and sparse CCA were fitted to the raw data, whereas multiCAsp and PCA+CCA used the same modality-specific PCA scores, with truncation dimensions chosen by a Kaiser-screened profile-likelihood rule~\cite{kaiser1960application,zhu2006automatic}. The comparison between multiCAsp and PCA+CCA therefore isolates the contribution of spatial modeling. We measured factor recovery by $R^2_{\mathrm{subspace}}$, the proportion of variation in the true factors explained by a regression on the estimated factors, which is invariant to rotations, permutations, and sign changes. We measured loading recovery by the mean absolute correlation between matched true and estimated raw-feature loading vectors, and spatial smoothness by the normalized graph-Laplacian energy $\hat s_c^\top L\hat s_c/\hat s_c^\top\hat s_c$ of each centered factor, where lower values indicate smoother variation over the spatial graph.
	
\begin{figure}[H]
		\centering
		\includegraphics[
		width=0.9\linewidth,
		height=0.34\textheight,
		keepaspectratio
		]{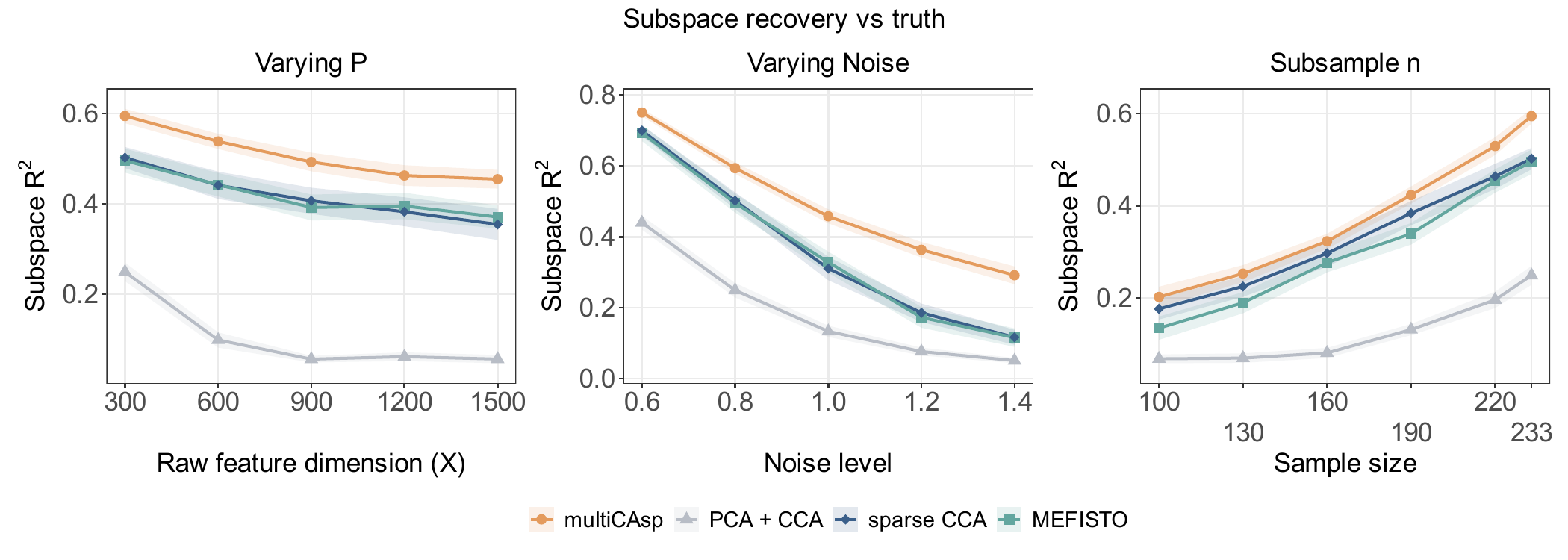}
		\caption{\textbf{Recovery of the shared latent spatial subspace.} Recovery is measured by $R^2_{\mathrm{subspace}}$ (higher is better). Panels vary $p_1$ (left), $\sigma$ (center), and $n$ (right) one at a time from their baseline values. PCA truncation dimensions were selected separately for each modality and replicate using Kaiser-screened profile likelihood on the log-transformed eigenvalue spectrum. Lines show means, and shaded regions show pointwise 95\% confidence intervals across 50 replicates.}
		\label{fig:subspace_recovery}
	\end{figure}

	\paragraph{Latent subspace recovery.}
	multiCAsp achieved consistently higher mean subspace recovery ($R^2_{\mathrm{subspace}}$) than PCA+CCA, sparse CCA, and MEFISTO across all simulation conditions (Figure~\ref{fig:subspace_recovery}).
	The performance gap relative to the next-best method was largest under high noise levels, suggesting a particular benefit of spatial regularization when
	shared signals are obscured by measurement noise. As raw feature dimensionality increased, multiCAsp
	maintained its advantage over the competing methods, with a particularly pronounced gap relative to PCA+CCA.
	Subspace recovery improved with increasing sample size for all methods, while multiCAsp consistently achieved the highest mean recovery.
	
\begin{figure}[!htbp]
		\centering
		\includegraphics[
		width=1.0\linewidth,
		height=0.49\textheight,
		keepaspectratio
		]{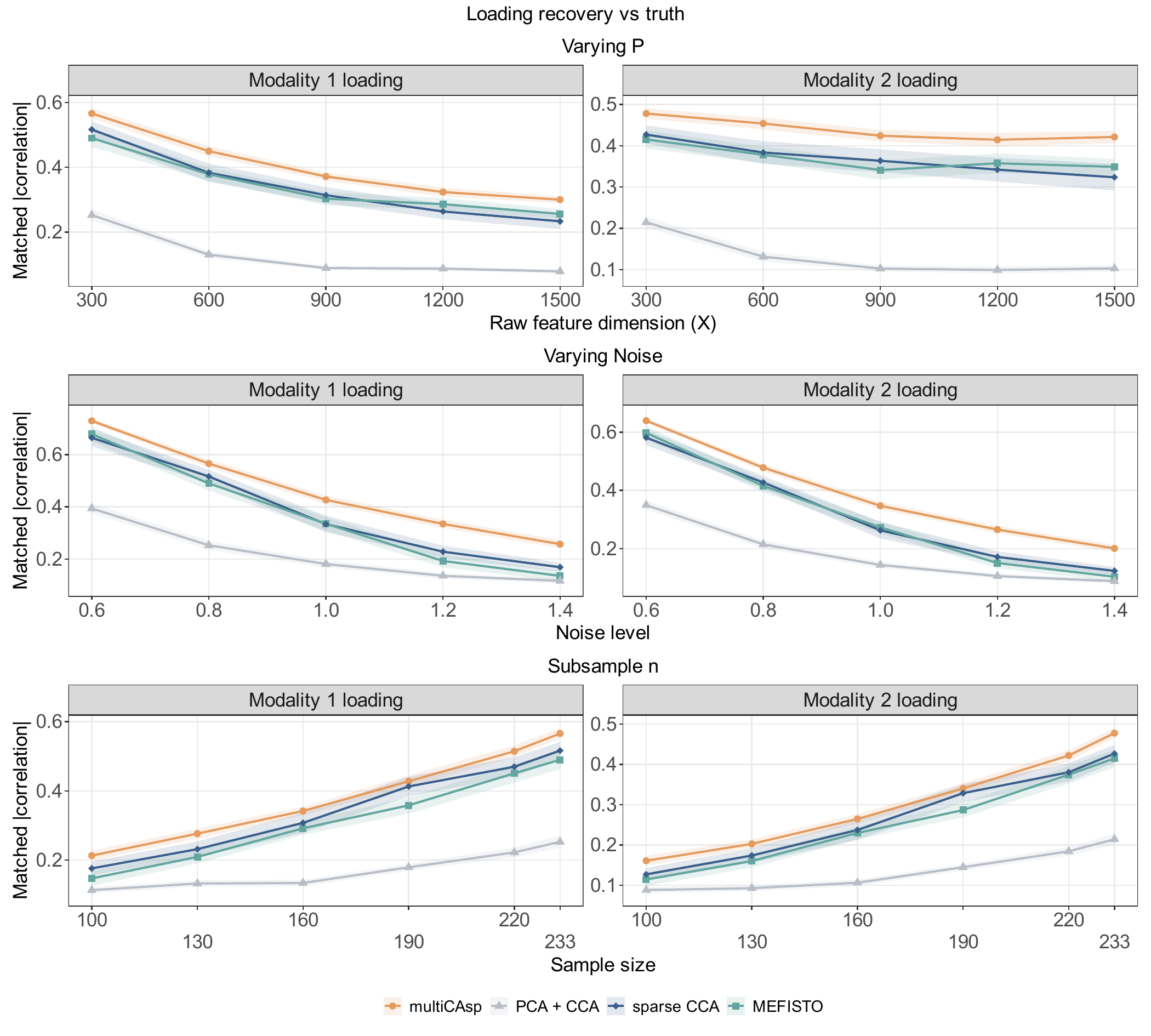}
		\caption{\textbf{Recovery of raw-feature loading patterns.} Recovery is measured by the mean absolute Pearson correlation
			between estimated and true raw-feature loading vectors (higher is better). Rows correspond to varying raw feature dimension $p_1$ (top), noise level $\sigma$
			(middle), and sample size $n$ (bottom). The left and right columns show loading recovery for modalities 1 and 2, respectively. Lines show means, and shaded regions indicate pointwise 95\% confidence intervals across
			50 replicates.}
		\label{fig:gram_low}
	\end{figure}

	\paragraph{Loading-structure recovery.}
	The score-level gains from multiCAsp were accompanied by recovery
	of the true raw-feature loading patterns in both modalities (Figure~\ref{fig:gram_low}). Across all simulation settings, multiCAsp achieved the highest mean absolute correlations between
	estimated and true loading vectors. Although loading recovery generally declined with increasing feature dimensionality and noise, multiCAsp maintained the highest mean recovery
	across both modalities, with more pronounced advantages under higher noise levels. Recovery improved with increasing sample size for all methods, while multiCAsp consistently retained the strongest performance across the full range of sample sizes.
	
\begin{figure}[!htbp]
		\centering
		\includegraphics[
		width=\linewidth,
		height=0.5\textheight,
		keepaspectratio
		]{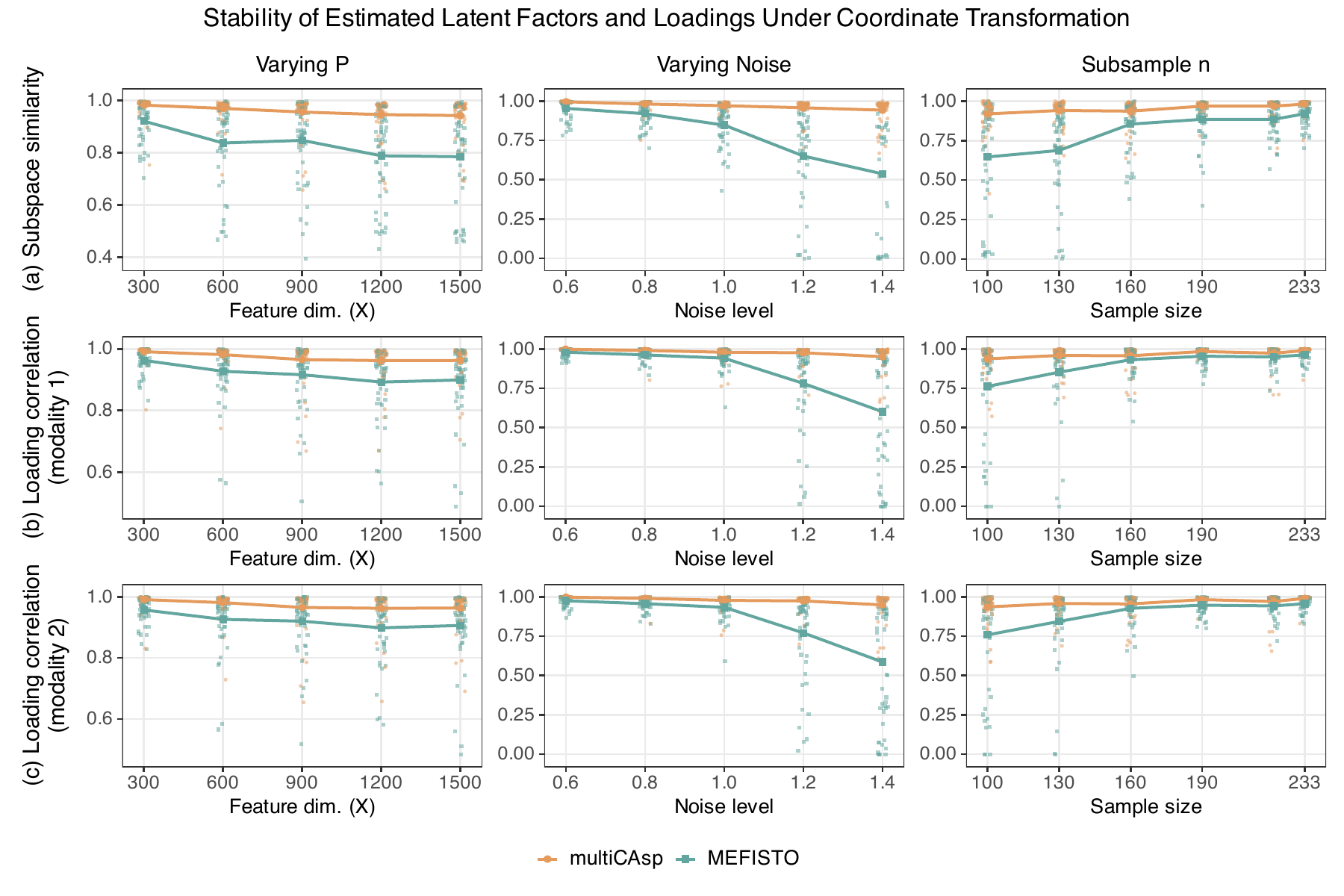}
		\caption{
			\textbf{Stability of estimated latent factors and feature loadings under coordinate transformation.} Columns correspond to varying raw feature dimension $p_1$ (left), noise level $\sigma$ (middle), and sample size $n$ (right). Rows show (a) latent subspace similarity, (b) loading correlations for modality 1, and (c) loading correlations for modality 2,
			comparing estimates obtained using original and transformed spatial coordinates. Points represent paired simulation replicates, and lines show mean stability across replicates. Higher values indicate greater stability.}
		\label{fig:warping_stability}
	\end{figure}

	\paragraph{Robustness to coordinate representation.} To mimic the distortion introduced when folded or irregular tissue is represented in two dimensions, we transformed the spatial coordinates by circle inversion~\cite{lin2018ordinarycircles} and evaluated the stability of estimated latent factors and feature loadings under this smooth nonlinear coordinate transformation. Across varying feature dimensions, noise levels, and sample sizes, multiCAsp maintained high similarity between estimates obtained using the original and transformed spatial coordinates (Figure~\ref{fig:warping_stability}). In comparison, MEFISTO exhibited greater variability, particularly under higher noise levels and smaller sample sizes. PCA+CCA and sparse CCA do not use spatial coordinates and are invariant to the transformation by construction, so they are not shown. These results support the advantage of multiCAsp's graph-based spatial prior, which makes it less sensitive than MEFISTO to how spatial relationships are represented and yields consistent latent factors and feature loadings under different coordinate representations. Details of the coordinate transformation are given in Appendix~\ref{app:sim}.
	
\begin{figure}[!htbp]
		\centering
		\includegraphics[
		width=\linewidth,
		height=0.34\textheight,
		keepaspectratio
		]{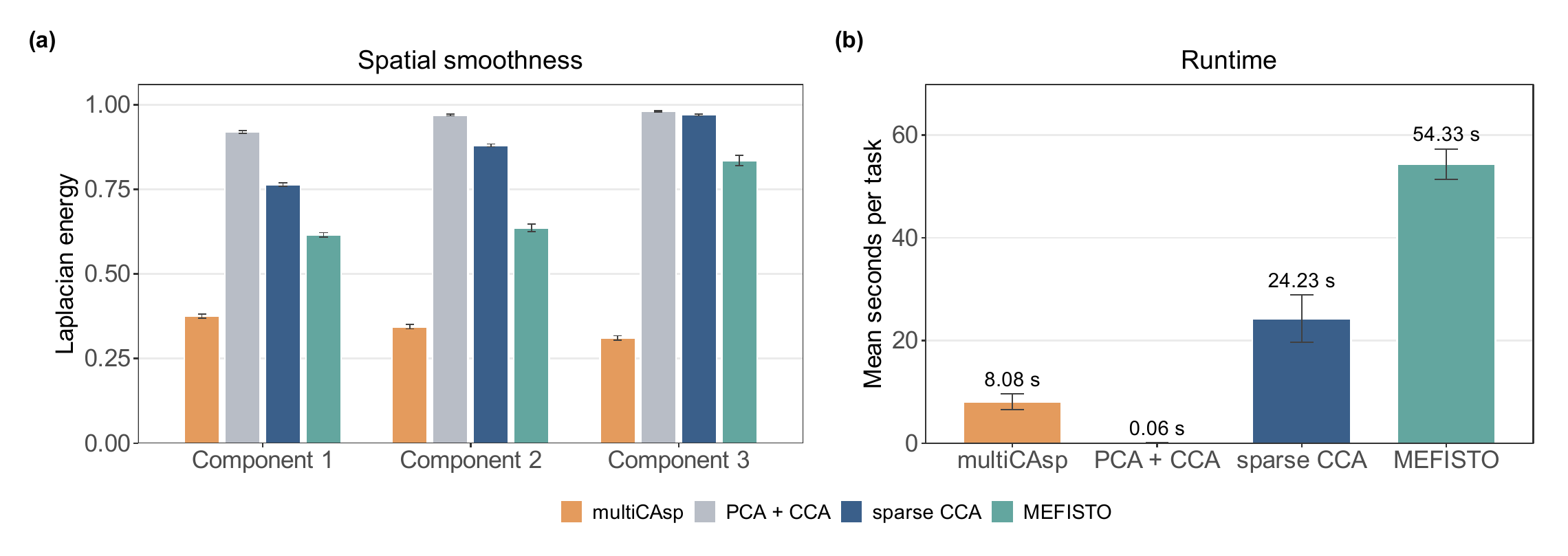}
		\caption{\textbf{Comparison of spatial smoothness and computational efficiency.} Results are summarized across all simulation settings varying feature dimension, noise level, and sample size, with 50 replicates per setting. Error bars indicate 95\% confidence intervals for the means. (a) Normalized graph-Laplacian energy of the recovered components, where lower values indicate smoother spatial patterns. (b) Mean runtime per simulation task.}
		\label{fig:spatial_smoothness}
	\end{figure}

	\paragraph{Spatial coherence and runtime.}
	multiCAsp consistently exhibited the lowest normalized graph-Laplacian energy across all recovered components among the four methods (Figure~\ref{fig:spatial_smoothness}a), indicating greater spatial smoothness with respect to the underlying tissue graph. In terms of computational efficiency, multiCAsp was considerably faster than sparse CCA and MEFISTO, while PCA+CCA
	remained the fastest method (Figure~\ref{fig:spatial_smoothness}b).
	
	Taken together, these results demonstrate that multiCAsp improves shared latent subspace and loading recovery across a range of simulation
	settings, with the largest relative gains under high noise, while producing spatially smoother factors at a lower computational cost than sparse CCA and MEFISTO.
	
	\subsection{Macaque brain multi-omic data}
	\label{sec:results-a26}
	
	The A26 dataset comprises paired targeted metabolomics (328 metabolites quantified by the Biocrates Q500 kit) and untargeted proteomics (9,990 protein groups by data-independent acquisition mass spectrometry) from 233 matched voxels ($\sim$4 mm per side) dissected from the right cortical hemisphere of an adult female rhesus macaque (\textit{Macaca mulatta})~\cite{wu2026integrated}. Processed proteomics data~\cite{wu2026proteomicsdata} and metabolomics data and sample-position tables~\cite{wu2026metabolomicsdata} are publicly archived. Spatial mapping of the samples is also described in detail in~\cite{wu2026integrated}. Metabolite and protein intensities were batch-corrected using ComBat~\cite{johnson2007adjusting} and outliers were removed by robust PCA-based distance screening prior to downstream analysis, using the preprocessing code archived in~\cite{wu2026preprocessingcode}. After preprocessing and standardization, PCA was performed separately for each modality, with truncation dimensions selected by the same rule as in simulations. The resulting PCA representations were used for both multiCAsp and PCA+CCA. Sparse CCA and MEFISTO were fitted directly to the unreduced modalities.
	
\begin{figure}[!htbp]
		\centering
		\includegraphics[
		width=\linewidth,
		height=0.8\textheight,
		keepaspectratio
		]{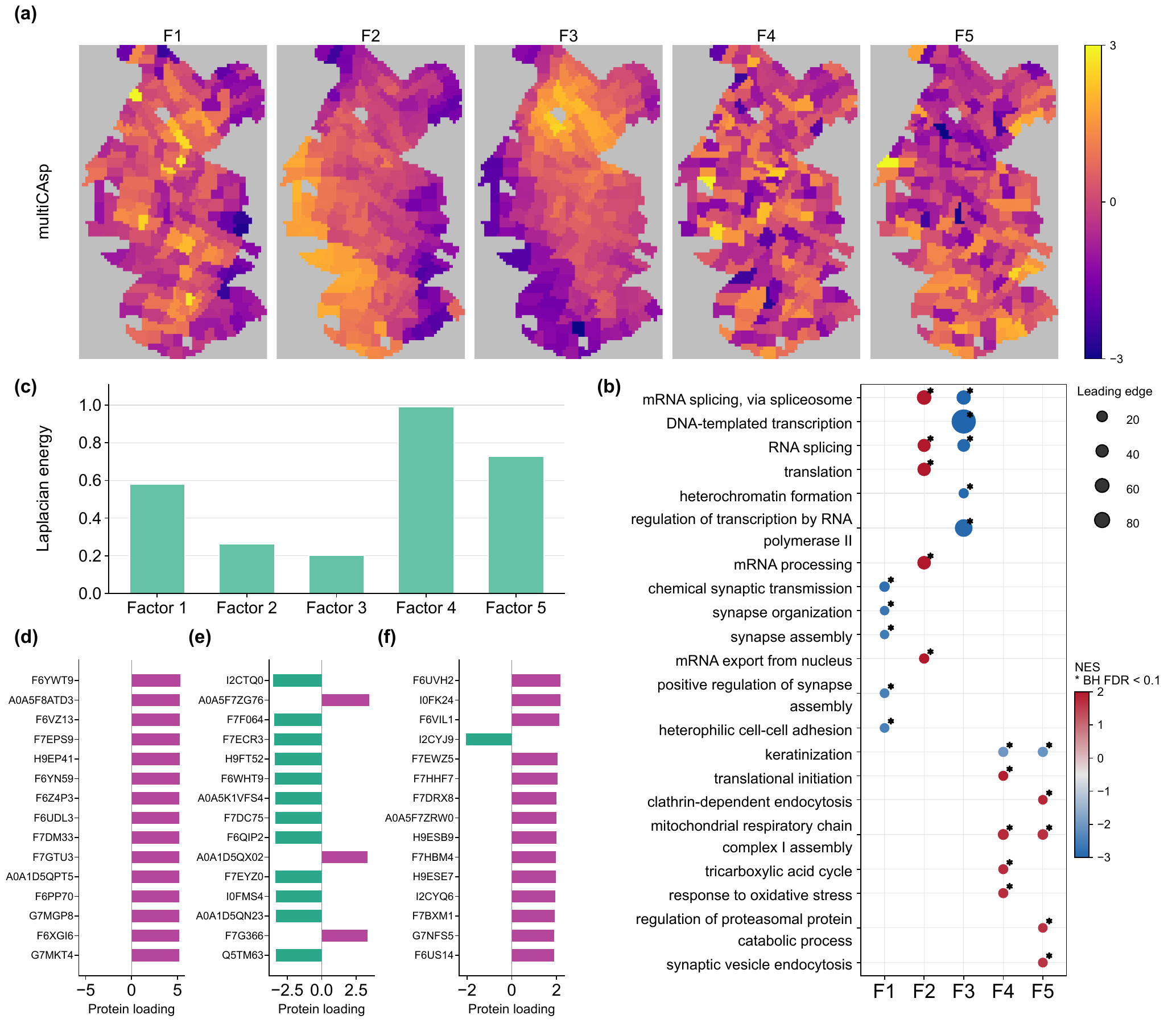}
		\caption{
			\textbf{Shared spatial factor analysis of the macaque brain multi-omic data.} (a) Spatial patterns of the five factors identified by multiCAsp, ordered by cross-modal agreement. (b) Gene set enrichment analysis (GSEA) of protein loadings across the five multiCAsp factors. Point size indicates the number of leading-edge genes, color indicates the normalized enrichment score (NES), and asterisks denote BH FDR $<0.1$. (c) Normalized graph-Laplacian energy for multiCAsp across the five factors. (d--f) Top protein loadings for Factor~1 (d), Factor~2 (e), and Factor~3 (f). }
		\label{fig:a26_heatmap}
	\end{figure}

	Figure~\ref{fig:a26_heatmap}a shows the shared latent factors from multiCAsp with rank $r=5$, ordered by cross-modal agreement (Section~\ref{sec:rotation}), and Figure~\ref{fig:a26_heatmap}c shows their normalized graph-Laplacian energy. Factors~2 and~3 showed the clearest regional organization and the lowest energy (0.26 and 0.20), whereas Factor~4 had energy close to 1, the expected value for spatially unstructured variation. Comparing each method's set of five factors, multiCAsp produced the smoothest factors of any method: Factors~2 and~3 (energy 0.26 and 0.20) were smoother than every factor from PCA+CCA, sparse CCA, and MEFISTO, whose smoothest factors had energy of about 0.34--0.36 (Supplementary Figure~\ref{fig:a26_methods}). The average energy across the five factors was lower for multiCAsp than for MEFISTO (about 0.55 versus 0.70) and comparable to that of PCA+CCA and sparse CCA.
	
	For each factor, we ranked features by their loadings and performed preranked gene set enrichment analysis (GSEA)~\cite{subramanian2005gene} against GO Biological Process gene sets, with Benjamini--Hochberg adjustment~\cite{benjamini1995controlling}. The spatially organized Factors~2 and~3 were both significantly enriched for RNA splicing, and were further distinguished by translation and mRNA processing for Factor~2 and transcriptional regulation for Factor~3 (Figure~\ref{fig:a26_heatmap}b). Among metabolites, Factor~3 featured more small molecules, such as lactate, succinate, and amino acids, compared to the other phospholipid-dominated factors (Supplementary Figure~\ref{fig:supp_a26_loadings}).
	
\begin{figure}[!htbp]
		\centering
		\includegraphics[
		width=\linewidth,
		height=0.33\textheight,
		keepaspectratio
		]{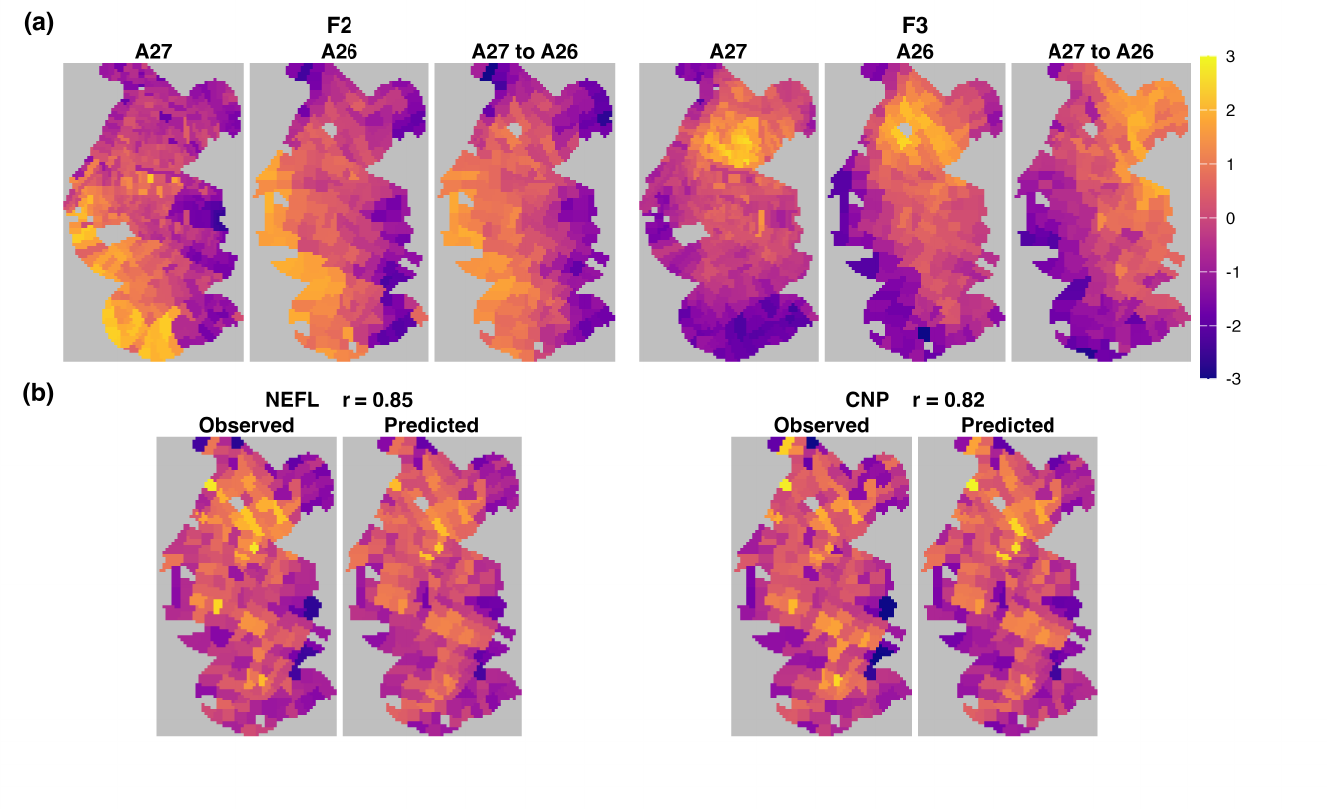}
		\caption{\textbf{Cross-animal transfer and cross-modal protein prediction.}
			(a) Spatial patterns of Factors 2 and 3. For each factor, the three columns show independently estimated factors in A27 and A26, followed by the transferred factor in A26. Transfer was performed using A26 metabolomic measurements and A27-derived metabolomic loadings. (b) Observed and predicted spatial abundances of NEFL and CNP in A26. Predictions were reconstructed using all five metabolomics-only transferred factors and  A27-derived proteomic loadings. Pearson correlations between observed and predicted abundances are shown.}
		\label{fig:a27_transfer}
	\end{figure}

	We further used a second macaque cortex dataset, A27, from a second adult rhesus macaque in the same study~\cite{wu2026integrated}, containing paired metabolomic and proteomic measurements, to evaluate cross-animal transferability. We fitted multiCAsp jointly to both modalities in A27 and held the resulting loadings and PCA projections fixed. We then estimated spatial factors in A26 using A26's spatial graph and its metabolomic measurements alone, and predicted A26 protein abundances from these factors and the A27 protein loadings. Observed A26 proteins were used only for evaluation, and the analysis was repeated in the reverse direction (Appendix~\ref{app:cross-animal}). For Factors~2 and~3, the transferred factors closely resembled those estimated independently in both animals (Figure~\ref{fig:a27_transfer}a), and this held in both transfer directions (Supplementary Figure~\ref{fig:multicasp_bidirectional_transfer}, which shows all five factors), indicating that shared spatial structures can be recovered across animals.
	
	We further evaluated cross-modal prediction using the transferred factors to reconstruct proteomic measurements in A26. The predicted spatial abundance patterns of NEFL and CNP closely resembled their observed distributions (Figure~\ref{fig:a27_transfer}b). The Pearson correlations were 0.85 for NEFL and 0.82 for CNP. These two proteins had among the strongest metabolite associations in the training animal, while prediction accuracy varied widely depending on the strength of this source association in both transfer directions (Supplementary Figure~\ref{fig:multicasp_coupling}). Complete results for all five factors in both transfer directions and additional protein-level prediction analyses are provided in Appendix~\ref{app:cross-animal}.
	
	\subsection{Gene--protein spatial multi-omic data}
	\label{sec:results-ge}
	
	The gene--protein dataset is a publicly available Xenium In Situ spatial multi-omic dataset from 10x Genomics~\cite{10xgenomics2025xenium}, profiling 465,534 cells in a Stage~III FFPE human renal cell carcinoma specimen using the Xenium Human Multi-Tissue and Cancer Panel (377 genes plus 28 custom genes) and a 27-plex protein panel. Cell-level measurements were aggregated into spatial bins. After excluding the bottom two grid rows, 697 bins were retained for multiCAsp analysis.
	
	For the higher-dimensional gene modality, PCA was performed after preprocessing, and the smallest number of principal components explaining at least 80\% of the total variance was retained ($K=56$). Protein features were standardized and entered directly without PCA. The resulting representations were used for both multiCAsp and PCA+CCA. The estimated factor subspace was stable for $K\geq 25$, with pairwise subspace similarities of 0.96--0.99 (Supplementary Figure~\ref{fig:ge_k_stability}, Appendix~\ref{sec:ge_k_sensitivity}), so the results are not sensitive to the exact choice of $K=56$. In contrast, the profile-likelihood rule used for the macaque data selected only $K=9$ here, which gave a noticeably different subspace (similarity 0.85--0.87). Sparse CCA was not applied to this dataset because its gene and protein features are highly collinear with dense signals, which conflicts with the sparsity assumption of sparse CCA.
	
\begin{figure}[!htbp]
		\centering
		\includegraphics[
		width=\linewidth,
		height=0.75\textheight,
		keepaspectratio
		]{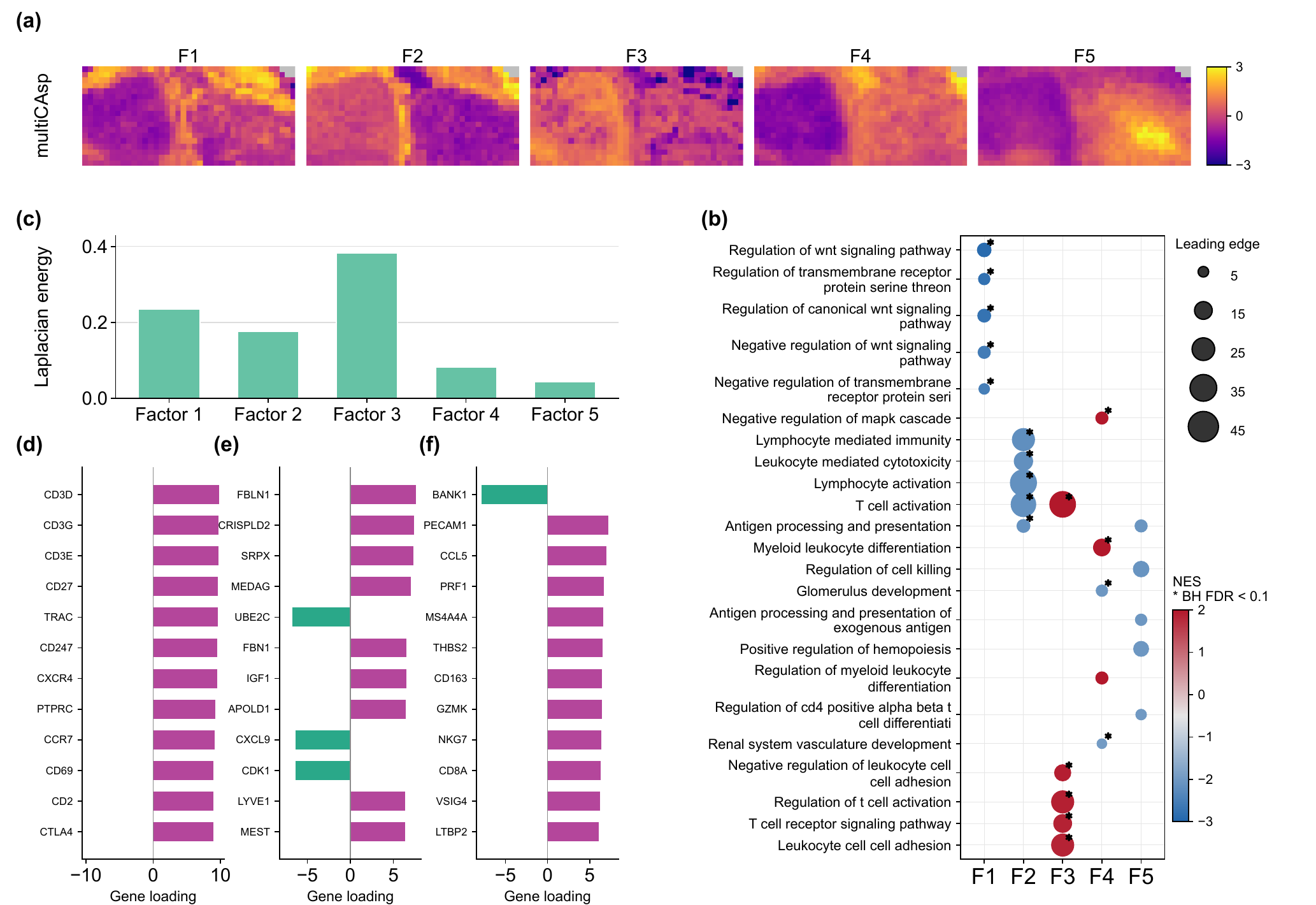}
		\caption{
			\textbf{Shared spatial factor analysis of the gene--protein multi-omic data.} (a) Spatial patterns of the five factors identified by multiCAsp, ordered by cross-modal agreement. (b) Gene set enrichment analysis (GSEA) of gene loadings across the five multiCAsp factors. Point size indicates the number of leading-edge genes, color indicates the normalized enrichment score (NES), and asterisks denote BH FDR $<0.1$. (c) Normalized graph-Laplacian energy for multiCAsp across the five factors. (d--f) Top gene loadings for Factor~1 (d), Factor~2 (e), and Factor~3 (f). }
		\label{fig:ge_heatmap}
	\end{figure}

	Figure~\ref{fig:ge_heatmap}a shows the shared latent factors from multiCAsp with rank $r=5$, ordered by cross-modal agreement, and Figure~\ref{fig:ge_heatmap}c shows their normalized graph-Laplacian energy. All five factors were spatially structured, with energy between 0.04 and 0.38, well below the value of about 1 expected for unstructured variation. Several factors were defined by the same cell-type markers in both modalities (Figure~\ref{fig:ge_heatmap}d--f and Supplementary Figure~\ref{fig:supp_ge_loadings}), and GSEA of the gene loadings corroborated these interpretations (Figure~\ref{fig:ge_heatmap}b). Factor~1 loaded on T-cell and leukocyte markers (genes CD3D, CD3E, and TRAC, proteins CD3E and CD45). Factor~3 loaded on cytotoxic T-cell markers (genes CD8A, GZMK, PRF1, and NKG7, protein CD8A) and was enriched for T-cell activation and T-cell receptor signaling. Factor~4 loaded on the macrophage marker CD163 in both modalities and was enriched for myeloid leukocyte differentiation. Factor~5 loaded on plasma-cell markers, including SDC1 and MZB1 among genes and CD138, the protein product of SDC1. Factor~2 contrasted stromal and vascular markers, such as FBLN1, alpha-SMA, CD31, and vimentin, with immune-activation markers, such as CXCL9, HLA-DR, PD-1, and LAG-3, and was enriched for lymphocyte activation, leukocyte-mediated cytotoxicity, and antigen presentation. This concordance between independently measured genes and proteins, supported at the pathway level, indicates that the shared factors capture cell-type programs detected by both assays.
	
	Comparing each method's set of five factors, the average graph-Laplacian energy was about half that of PCA+CCA and MEFISTO (0.18 versus 0.38 and 0.39), and the smoothest multiCAsp factors (energy 0.04 and 0.08) were smoother than every factor from either comparison method, whose smoothest factors had energy of about 0.15--0.16 (Supplementary Figure~\ref{fig:ge_methods}).
	
	Taken together, the two real-data applications show that multiCAsp recovers spatially coherent shared latent factors with distinct feature-level signatures for downstream biological interpretation.
	
	\section{Discussion}
	\label{sec:discussion}
	
	We introduced multiCAsp, a probabilistic framework for identifying shared spatial structure across multiple omic modalities. Its defining feature is that spatial dependence is encoded through a neighborhood graph rather than through distances between coordinates. This choice has three consequences. First, the prior follows tissue adjacency, so it accommodates three-dimensional and irregular tissues, samples that are close in space but not adjacent in the tissue, and multiple tissue sections, and the graph can be curated when coordinates are unreliable. Second, the results depend less on how coordinates are assigned, as shown by the greater stability of multiCAsp than MEFISTO under a nonlinear coordinate transformation. Third, the prior shares its eigendecomposition with the graph Laplacian, which makes estimation, including that of the spatial smoothing parameter, efficient, and multiCAsp ran substantially faster than MEFISTO. In simulations, multiCAsp also improved latent subspace and feature-loading recovery compared with competing methods, with particularly strong gains under high noise. In two real datasets with different spatial designs, multiCAsp recovered spatially coherent shared factors with interpretable feature loadings, and its most spatially organized factors were smoother than any factor from the comparison methods. Factors and loadings learned in one specimen also transferred to another and supported prediction of one modality from another.

	Several aspects of multiCAsp merit further investigation. First, multiCAsp does not perform variable selection. Its feature loadings are dense, in part because the PCA bases combine many features, so interpretation relies on ranking features by loading magnitude rather than on a sparse set of selected features. Second, the framework requires selecting the latent dimension and modality-specific PCA truncation levels, which may influence the recovered factors and their interpretation. Finally, our real-data evaluations cover a limited number of datasets and assay combinations, and further validation across tissues, spatial resolutions, and biological settings will help establish generalizability.
	
	Future work could incorporate sparsity-inducing priors or penalties on the feature loadings to enable variable selection, and develop methods for selecting the latent dimension and PCA truncation levels. Developing uncertainty quantification for shared factors and feature loadings would also strengthen statistical inference. Although the model and estimation algorithm apply to any number of modalities, spatially resolved datasets with three or more matched modalities remain scarce. As emerging platforms make such data and multi-section datasets more widely available, applications to them will further demonstrate the flexibility of the multiCAsp framework. Overall, multiCAsp provides a spatially informed approach to multi-omic integration and offers a foundation for studying shared molecular variation in complex biological tissues.
	
	

	
	\bibliographystyle{unsrtnat}
	\bibliography{ref_multiCAsp}
	
	
	\appendix
	
	\par\bigskip
	\noindent\rule{\textwidth}{0.5pt}
	\par\medskip

	\section*{Appendix: Supplementary Analyses}
	
	\section{Simulation details}\label{app:sim}
	
	\paragraph{Shared spatial latent factors.}
	We used the normalized graph Laplacian $L$ of the macaque A26 data (Section~\ref{sec:results-a26}, $n = 233$ spatial locations), with eigendecomposition $L = QVQ^\top$.  The smoothness parameter was set to \[ \eta_{\mathrm{true}} = \frac{50}{\operatorname{median}\{\operatorname{diag}(V)\}}, \] and $r = 3$ shared spatial latent components were generated as $s_k \sim \mathcal{N}_n(0, W(\eta_{\mathrm{true}}))$, $k = 1, 2, 3$. All methods were fitted with $r=3$ in the simulation.
	
	For the feature-dimensionality and noise-level simulation settings, the spatial covariance matrix $W(\eta_{\mathrm{true}})$ was held fixed, while the shared latent factors were independently generated for each
	replicate.
	
	\paragraph{Modality-specific low-dimensional signals.}
	For both simulated modalities we set $d_1 = d_2 = 10$ and generated dense loading matrices $A_0^{(m)} \in \mathbb{R}^{r \times d_m}$ with orthonormal rows. Component-specific signal strengths were introduced through \[ A_{\mathrm{low,true}}^{(m)} = \Lambda^{(m)} A_0^{(m)}, \qquad \Lambda^{(1)} = \Lambda^{(2)} = \operatorname{diag}(8, 6, 4). \]
	
	\paragraph{High-dimensional collinear expansion.}
	Low-dimensional signals were embedded into high-dimensional raw feature spaces via expansion matrices $P_{\mathrm{true}}^{(m)} \in \mathbb{R}^{p_m \times d_m}$, constructed by organizing features into $B = 5$ matched nonoverlapping blocks. Within each block, a Gaussian random matrix was orthogonalized by QR decomposition and inserted into the corresponding submatrix of $P_{\mathrm{true}}^{(m)}$. Entries with absolute magnitude below $0.02$ were thresholded to zero, and each column was renormalized to unit $\ell_2$ norm.
	This construction produced approximately orthogonal expansion matrices with localized, block-structured loadings, introducing dependence among raw features through shared low-dimensional signals.
	
	\paragraph{Observed data.}
	The two observed data matrices were generated as \[ X^{(m)} = S_{\mathrm{true}} A_{\mathrm{raw,true}}^{(m)} + \sigma_m E^{(m)}, \qquad E^{(m)} \sim \mathcal{N}_{n,p_m}(0, I_n, I_{p_m}), \] where $A_{\mathrm{raw,true}}^{(m)} = A_{\mathrm{low,true}}^{(m)} P_{\mathrm{true}}^{(m)\top}$. Unless otherwise specified, we set $\sigma_1 = \sigma_2 = \sigma$. The baseline settings were $n = 233$, $p_1 = 300$, $p_2 = 600$, and $\sigma = 0.8$. We varied three aspects of the data-generating mechanism one at a time:
	\begin{itemize}
		\item \textit{Feature dimension:} $p_1 \in \{300, 600, 900, 1200, 1500\}$,
		with $p_2 = 600$ fixed.
		\item \textit{Noise level:} $\sigma \in \{0.6, 0.8, 1.0, 1.2, 1.4\}$.
		\item \textit{Sample size:} $n \in \{100, 130, 160, 190, 220, 233\}$, obtained by sampling spatial locations without replacement from the full $n = 233$ grid. For each subsample, $L_n$ was defined as the principal submatrix of the full-grid Laplacian restricted to the retained locations. The spatial smoothing parameter and covariance matrix were then recalculated
		using $L_n$ following the same procedure described above.
	\end{itemize}
	Each setting was evaluated using 50 independently generated datasets. For multiCAsp, the Delaunay triangulation was rebuilt for each subsampling. Results were summarized using the replicate mean and 95\% confidence intervals for the mean.
	
	\paragraph{PCA dimension selection.}
	PCA was performed separately for the two simulated modalities after centering and scaling each feature to unit variance. Because the features were standardized, principal components with
	eigenvalues less than or equal to one were first excluded following the
	Kaiser criterion~\cite{kaiser1960application}.
	Among the remaining components, the truncation dimension was selected using
	a modified profile-likelihood criterion based on Zhu and Ghodsi
	~\cite{zhu2006automatic}, applied to the log-transformed eigenvalue spectrum. For each candidate truncation point, the ordered log-eigenvalues were partitioned into two Gaussian groups with distinct means and a common variance, and the truncation point maximizing the profile likelihood was selected. The resulting dimensions $K_1$ and $K_2$ were selected independently for each modality and simulation replicate, with a minimum of three components and an upper bound of
	$\min(n-1,p_m)$.
	
	\paragraph{Coordinate transformation.}
	To mimic geometric distortion that can arise when folded or irregular
	anatomical structures are represented in a two-dimensional coordinate
	system, we applied a nonlinear transformation based on circle
	inversion~\cite{lin2018ordinarycircles}.
	Let $\mathbf{x}_i= (x_{i1},x_{i2})^\top\in\mathbb{R}^2$ denote the original coordinate of spatial location $i$, and let
	$\mathbf{c}=(c_1,c_2)^\top$ denote a fixed inversion center. Using unit-circle inversion centered at $c$, the intermediate transformed coordinate was defined as \[\mathbf{x}_i^{\mathrm{inv}}
	=\frac{\mathbf{x}_i-\mathbf{c}}
	{\|\mathbf{x}_i-\mathbf{c}\|_2^2}.\]
	
	To preserve the overall spatial scale, the inverted coordinates were
	isotropically rescaled so that their median pairwise Euclidean distance
	matched that of the original coordinates. Specifically, we defined \[\gamma = \frac{\operatorname{median}_{i<j}
		\|\mathbf{x}_i-\mathbf{x}_j\|_2}{
		\operatorname{median}_{i<j}
		\|\mathbf{x}_i^{\mathrm{inv}}
		-\mathbf{x}_j^{\mathrm{inv}}\|_2},\]
	and used $\mathbf{x}_i^{\mathrm{transformed}}=\gamma\,\mathbf{x}_i^{\mathrm{inv}}$. The inversion center was chosen at grid location M24, an unobserved location near the sampled cortical region, allowing a localized nonlinear distortion to be introduced without placing the inversion center on an observed sample.
	
	The same simulated datasets and true latent factors were analyzed by all methods under this scenario, with only the spatial coordinates altered. For multiCAsp, the spatial neighborhood graph and normalized graph Laplacian were reconstructed from the transformed coordinates, and the corresponding spatial covariance matrix was used for model fitting. For MEFISTO, the transformed Euclidean coordinates were used in its Gaussian process prior. PCA+CCA and sparse CCA do not use spatial coordinates directly. For multiCAsp, the Delaunay triangulation was rebuilt for each coordinate transformation.

	\paragraph{Methods compared.}
	We compared four methods:
	\begin{enumerate}[label=(\roman*)]
		\item
		\textbf{multiCAsp}: multiCAsp was fitted to the modality-specific PCA
		score matrices using the adaptively selected dimensions $K_1$ and $K_2$.
		\item
		\textbf{MEFISTO}: MEFISTO was fitted directly to the raw data matrices
		$X^{(1)}$ and $X^{(2)}$, with spatial
		coordinates incorporated through Gaussian process priors.
		\item \textbf{PCA+CCA}: standard CCA was fitted to the same
		PCA score matrices used by multiCAsp, providing a nonspatial comparison
		under identical dimension reduction.
		\item \textbf{Sparse CCA}: sparse CCA was fitted directly to the raw data
		matrices $X^{(1)}$ and $X^{(2)}$, with sparsity parameters selected using the permutation-based procedure in \texttt{PMA::CCA.permute}.
	\end{enumerate}
	
	\paragraph{Evaluation metrics.}
	For simulation evaluation, we used the latent factor and loading estimates obtained at EM convergence, prior to the whitening and rotation described in Section~\ref{sec:rotation}. 
	
	\textit{Latent subspace recovery:}
	Let $S_{\mathrm{true}} \in \mathbb{R}^{n \times r}$ and $\hat{S} \in \mathbb{R}^{n \times r}$ denote the true and estimated shared spatial latent factors, respectively.
	We centered both matrices column-wise to obtain $S_c$ and $\hat{S}_c$, and regressed $S_c$ on $\hat{S}_c$ using multivariate least squares with an intercept.
	Denoting the fitted values by $\widehat{S}_{\mathrm{true},c}$, we defined \[R_{\mathrm{subspace}}^2 = 1 - \frac{\|S_c -\widehat{S}_{\mathrm{true},c}\|_F^2}{\|S_c\|_F^2}.\] This metric measures the proportion of centered variation in the true latent factors explained by the estimated latent subspace. It is invariant to rotations, permutations, and sign changes of the estimated latent components.
	
	\textit{Loading-structure recovery:}
	Loading recovery was evaluated in the original feature space. For modality
	$m$, the true raw-feature loading matrix was \[A_{\mathrm{raw,true}}^{(m)}
	=A_{\mathrm{low,true}}^{(m)} P_{\mathrm{true}}^{(m)\top} \in \mathbb{R}^{r \times p_m}.\]
	For multiCAsp, the estimated reduced-space loadings were mapped back to the original feature space as
	\[\hat A_{\mathrm{raw}}^{(m)}=\hat A^{(m)}\hat P^{(m)\top}.\]
	
	The PCA+CCA loadings were mapped back to the original feature space in the
	same way. Sparse CCA was fitted directly to the raw features, and MEFISTO feature weights were extracted in the original feature space, so their
	corresponding raw-feature loadings were used directly.
	
	For each modality, estimated loading vectors were matched to the true loading vectors using the Hungarian algorithm to maximize the sum of absolute Pearson correlations. This matching was performed independently of the factor-score matching used for spatial smoothness.
	
	Loading recovery was then summarized by the mean absolute Pearson correlation between the matched estimated and true raw-feature loading vectors:
	\[
	R_{\mathrm{loading}}^{(m)}=\frac{1}{r}
	\sum_{k=1}^{r} \left|\operatorname{corr}
	\left(a_{\mathrm{raw,true},k}^{(m)},
	\hat{a}_{\mathrm{raw},k}^{(m)}\right)
	\right|.
	\]
	Higher values indicate more accurate recovery of the raw-feature loading patterns.

	\textit{Graph-Laplacian spatial roughness:} For each estimated component $\hat{s}_k$, after removing its empirical mean, spatial roughness was quantified by the normalized graph-Laplacian energy evaluated on the original spatial graph: \[ \mathcal{E}_L(\hat{s}_k) = \frac{\hat{s}_{k,c}^\top L\, \hat{s}_{k,c}} {\hat{s}_{k,c}^\top \hat{s}_{k,c}}. \] Lower values indicate smoother variation over the spatial graph.
	
	For factor-wise comparisons, estimated factors were matched to the true factors using the Hungarian algorithm based on absolute correlations between factor scores.
	
	\textit{Stability under coordinate transformation:}
	We evaluated the stability of estimated latent factors and feature loadings by directly comparing estimates under original and transformed spatial coordinates. Latent subspace similarity was measured by the squared Frobenius norm of the cross-product between orthonormal
	bases of the centered estimated factor matrices, normalized by the smaller number of retained factors.
	Feature loading stability was measured by the mean absolute Pearson correlation between matched raw-feature loading vectors estimated under the two coordinate conditions, using the retained factors for each method. Higher values indicate greater stability of the estimated latent factors and loadings.
	
	\section{Sensitivity to gene-expression PCA truncation}
	\label{sec:ge_k_sensitivity}
	
	Because PCA truncation is performed before fitting multiCAsp, we examined
	the sensitivity of the gene--protein analysis to the number of retained gene-expression principal components. The protein modality was retained in full, the latent dimension was fixed at $r=5$, and the spatial regularization parameter $\eta$ was re-estimated for each value of $K$.
	
	For each pair of truncation dimensions $K_i$ and $K_j$, we compared the
	estimated multiCAsp score subspaces using
	\[
	\operatorname{Sim}(K_i,K_j)
	=
	1-
	\frac{
		\left\|\mathcal{P}_{K_i}-\mathcal{P}_{K_j}\right\|_F
	}{
		\sqrt{2r}
	},
	\]
	where $\mathcal{P}_K$ denotes the orthogonal projection matrix onto the
	$r$-dimensional score subspace estimated using $K$ gene-expression
	principal components.

	\section{Cross-animal transfer and cross-modal prediction}
	\label{app:cross-animal}
	
	For the macaque data, modalities 1 and 2 denote
	proteomics and metabolomics, respectively.
	Using A27 as the training animal, we fitted
	multiCAsp jointly to both modalities.
	This yielded the loading matrices
	$\hat A_{27}^{(1)}$ and $\hat A_{27}^{(2)}$,
	along with the PCA projection matrices
	$P_{27}^{(1)}$ and $P_{27}^{(2)}$.
	These matrices were held fixed during transfer
	to the validation animal, A26.
	
	The A26 metabolomic data were aligned and
	projected into the A27 PCA space using
	the training-derived preprocessing parameters.
	Using only these metabolomic measurements,
	we estimated the A26 spatial factors through
	
	\[
	\widetilde X_{26}^{(2)}
	\approx
	S_{26}^{(2)}\hat A_{27}^{(2)},
	\]
	
	where $\widetilde X_{26}^{(2)}$ denotes the
	A26 metabolomic data in the A27 PCA space,
	and $S_{26}^{(2)}$ denotes the latent factors
	inferred from metabolomics alone.
	Estimation incorporated the A26 spatial graph,
	with the spatial parameter and metabolomic noise
	variance updated iteratively.
	
	The estimated spatial factors were then used
	to predict the A26 proteomic measurements:
	
	\[
	\widehat X_{26}^{(1)}
	=
	\hat S_{26}^{(2)}
	\hat A_{27}^{(1)}
	P_{27}^{(1)\top}.
	\]
	
	The resulting predictions were expressed on
	the A27-standardized feature scale.
	Observed A26 proteomic measurements were
	used only to evaluate prediction accuracy.
	The analysis was repeated in the reverse
	direction, using A26 for training and
	A27 for validation.

	We further examined the relationship between source cross-modal associations and cross-animal
	protein prediction performance
	(Figure~\ref{fig:multicasp_coupling}). For each protein $j$, we fitted a linear regression using metabolomic principal component scores
	in the source animal:
	
	\begin{equation}
	X_{\mathrm{src},\cdot j}=\beta_{0j}
	+Z_{Y,\mathrm{src}}\boldsymbol{\beta}_j+\varepsilon_j,
	\end{equation}
	
	where $Z_{Y,\mathrm{src}}$ denotes the metabolomic principal component scores. Source cross-modal
	association was measured by the multiple correlation coefficient:
	
	\begin{equation}
	R_j^{\mathrm{source}}=
	\operatorname{corr}\left(X_{\mathrm{src},\cdot j},\widehat X_{\mathrm{src},\cdot j}^{\mathrm{OLS}}\right).
	\end{equation}
	
	Cross-animal prediction performance was measured by the Pearson correlation between observed
	and predicted protein abundances in the target animal:
	
	\begin{equation}
	r_j^{\mathrm{transfer}} =\operatorname{corr}\left(
	X_{\mathrm{tgt},\cdot j},
	\widehat X_{\mathrm{tgt},\cdot j}^{\mathrm{multiCAsp}}
	\right).
	\end{equation}

	
	
	\clearpage

	\clearpage
	
	\setlength{\floatsep}{7pt plus 2pt minus 2pt}
	\setlength{\textfloatsep}{7pt plus 2pt minus 2pt}
	\setlength{\intextsep}{7pt plus 2pt minus 2pt}
	
	\renewcommand{\topfraction}{0.95}
	\renewcommand{\bottomfraction}{0.95}
	\renewcommand{\textfraction}{0.05}
	\renewcommand{\floatpagefraction}{0.75}
	\setcounter{topnumber}{4}
	\setcounter{bottomnumber}{4}
	\setcounter{totalnumber}{6}
	
	\captionsetup[figure]{
		font={small,stretch=1},
		skip=4pt,
		justification=raggedright,
		singlelinecheck=false
	}
	
	\section*{Main Figures}
	
	\clearpage
	
	\setcounter{figure}{0}
	\renewcommand{\thefigure}{S\arabic{figure}}
	\renewcommand{\theHfigure}{S\arabic{figure}}
	
	\setcounter{table}{0}
	\renewcommand{\thetable}{S\arabic{table}}
	\renewcommand{\theHtable}{S\arabic{table}}
	
	\section*{Supplementary Figures}
	
	\begin{figure}[H]
		\centering
		\includegraphics[
		width=\linewidth,
		height=0.75\textheight,
		keepaspectratio
		]{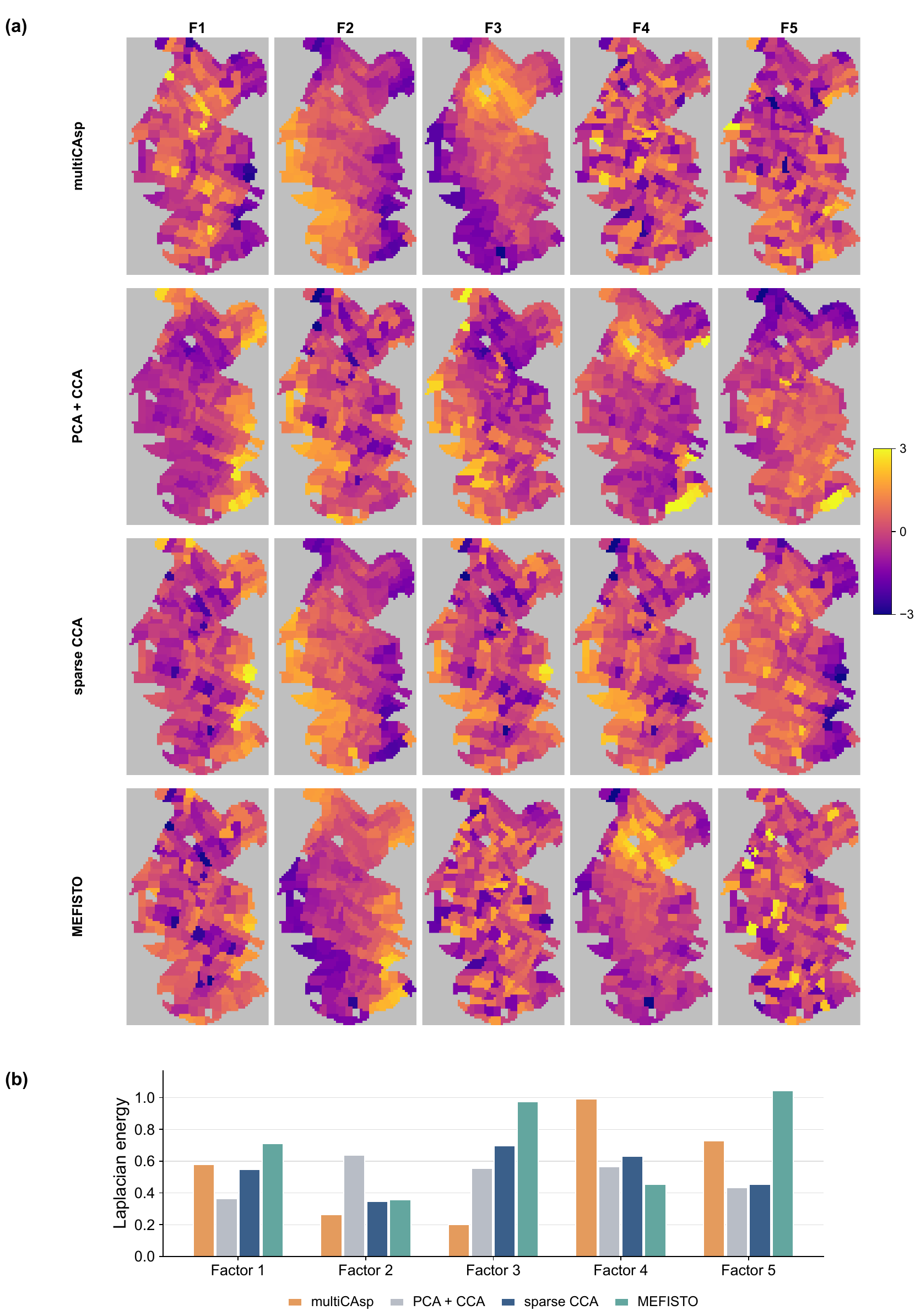}
		\caption{
			\textbf{Comparison of spatial factors across methods in the macaque brain multi-omic data.}
			Panel (a) shows spatial score maps for five factors recovered by multiCAsp,
			PCA+CCA, sparse CCA, and MEFISTO.
			Panel (b) summarizes spatial smoothness using normalized graph
			Laplacian energy, where lower values indicate more spatially coherent factor
			patterns. Spatial factors are identifiable up to a sign flip. 
		}
		\label{fig:a26_methods}
	\end{figure}
	
	\begin{figure}[!htbp]
		\centering
		\includegraphics[
		width=\linewidth,
		height=0.7\textheight,
		keepaspectratio
		]{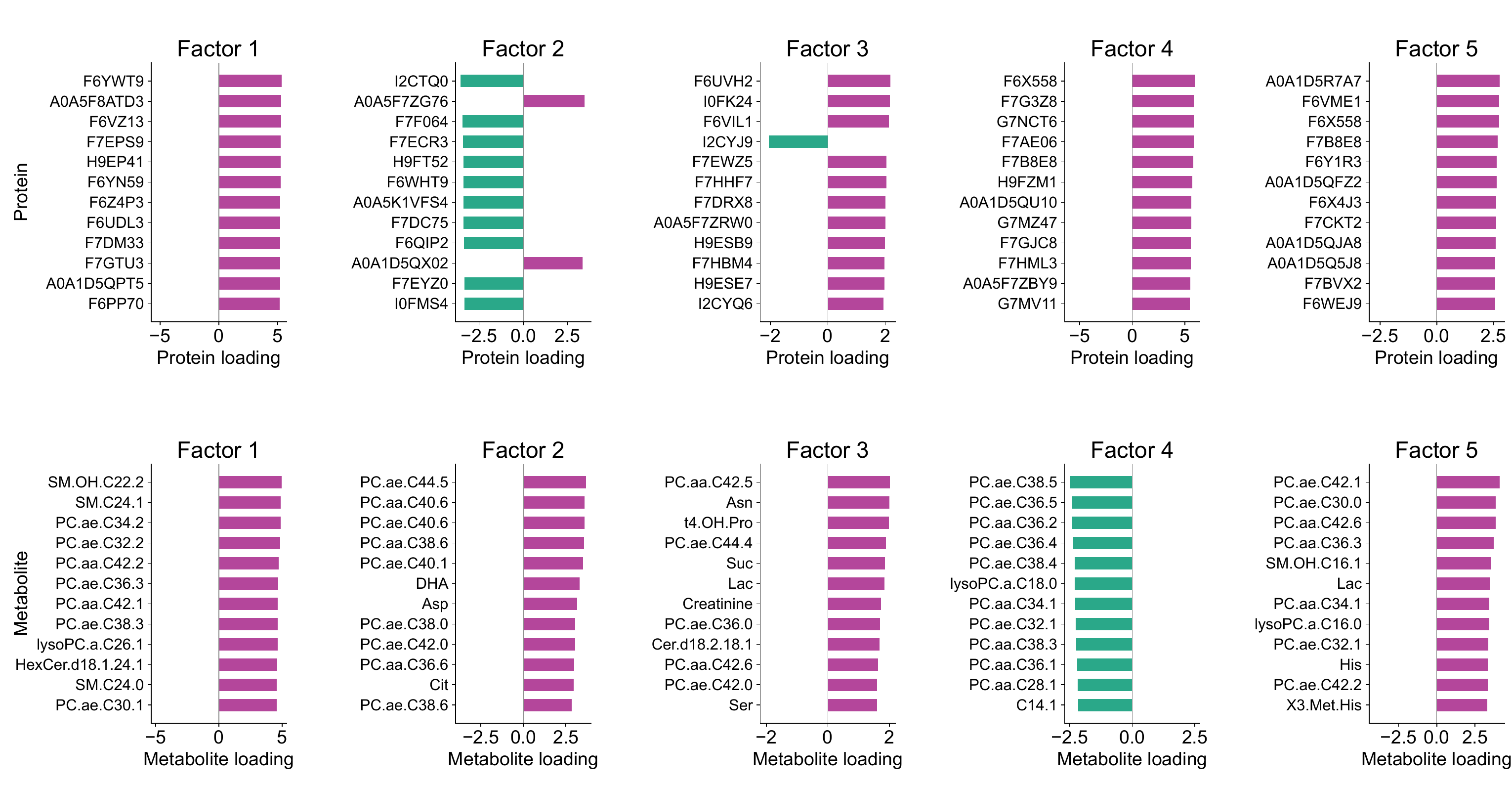}
		\caption{\textbf{Feature-loading patterns of shared spatial factors in the macaque brain multi-omic dataset.}
			The upper row shows protein loadings, and the lower
			row shows metabolite loadings. For each factor and modality, the 12 features with the largest absolute loading values are displayed. The loading profiles illustrate distinct contributions
			from both molecular modalities to the shared
			spatial factors.
		}
		\label{fig:supp_a26_loadings}
	\end{figure}
	
	\begin{figure}[!htbp]
		\centering
		\includegraphics[
		width=0.94\linewidth,
		height=0.40\textheight,
		keepaspectratio
		]{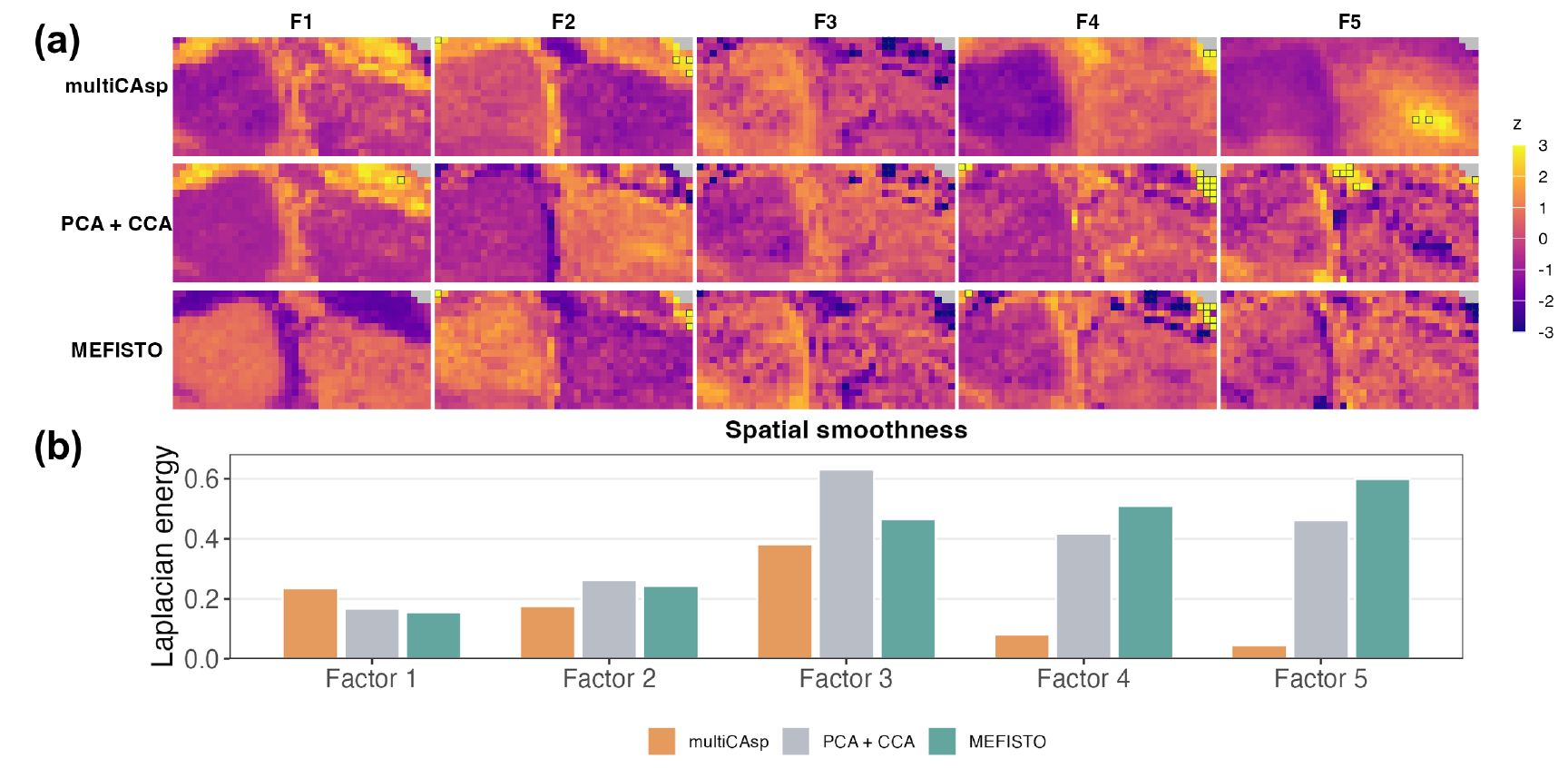}
		\caption{
			\textbf{Comparison of spatial factors across methods in the spatial
				gene--protein dataset.}
			Panel (a) shows spatial score maps for five factors recovered by multiCAsp, PCA+CCA, and MEFISTO.
			Panel (b) summarizes spatial smoothness using normalized graph
			Laplacian energy, where lower values indicate more spatially coherent factor
			patterns. Spatial factors are identifiable up to a sign flip.
		}
		\label{fig:ge_methods}
	\end{figure}
	
	\begin{figure}[!htbp]
		\centering
		\includegraphics[
		width=\linewidth,
		height=0.39\textheight,
		keepaspectratio
		]{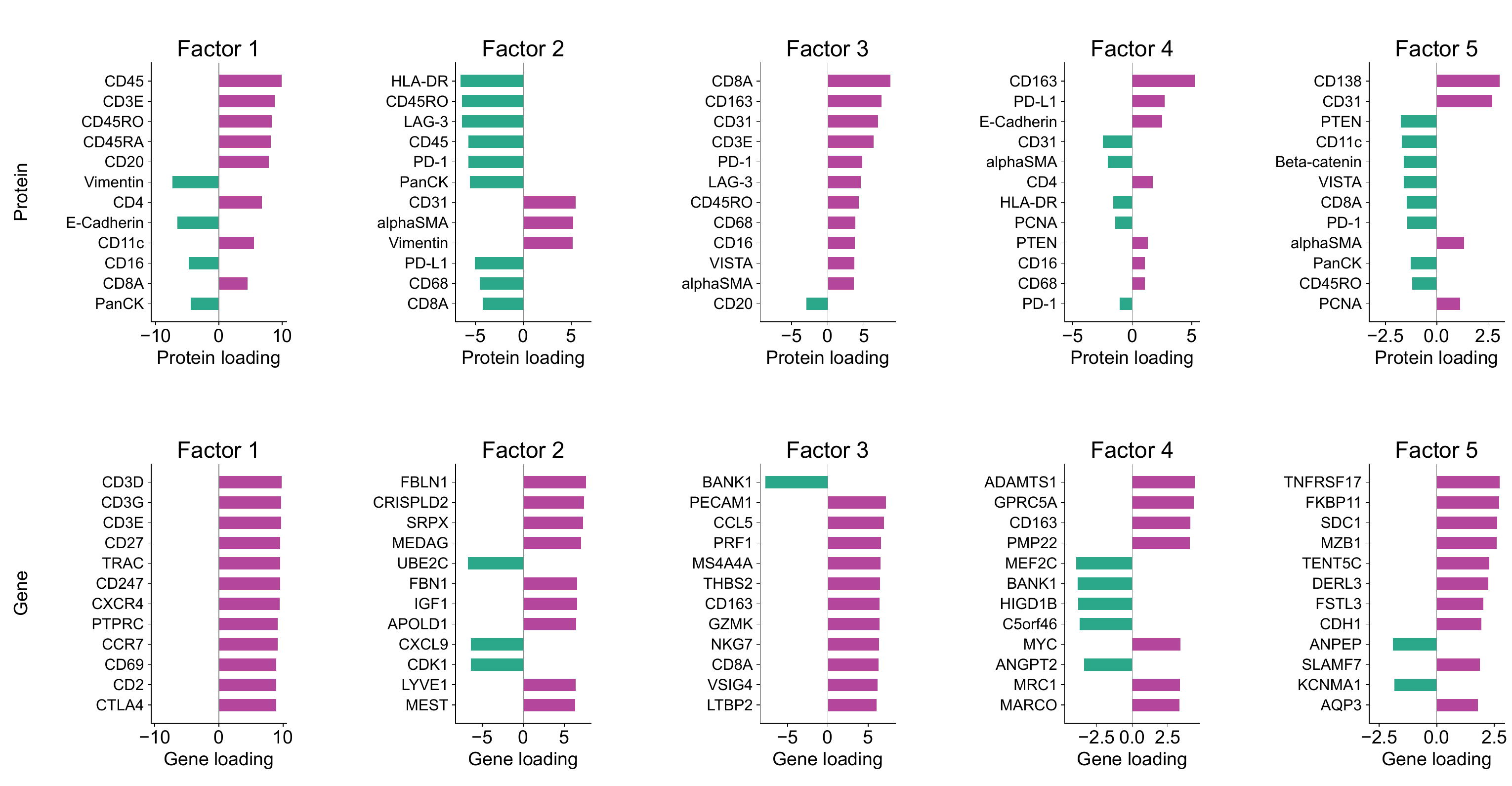}
		\caption{\textbf{Feature-loading patterns of shared spatial factors in the gene--protein dataset.} The upper row shows protein loadings, and the lower row shows gene loadings. For each factor and modality, the 12 features with the largest absolute loading values are displayed.
			Several factors are defined by the same cell-type markers in both modalities, including T-cell markers in Factor~1, cytotoxic T-cell markers in Factor~3, CD163 in Factor~4, and plasma-cell markers (SDC1 and its protein product CD138) in Factor~5.
		}
		\label{fig:supp_ge_loadings}
	\end{figure}
	
	\begin{figure}[!htbp]
		\centering
		\includegraphics[
		width=\linewidth,
		height=0.6\textheight,
		keepaspectratio
		]{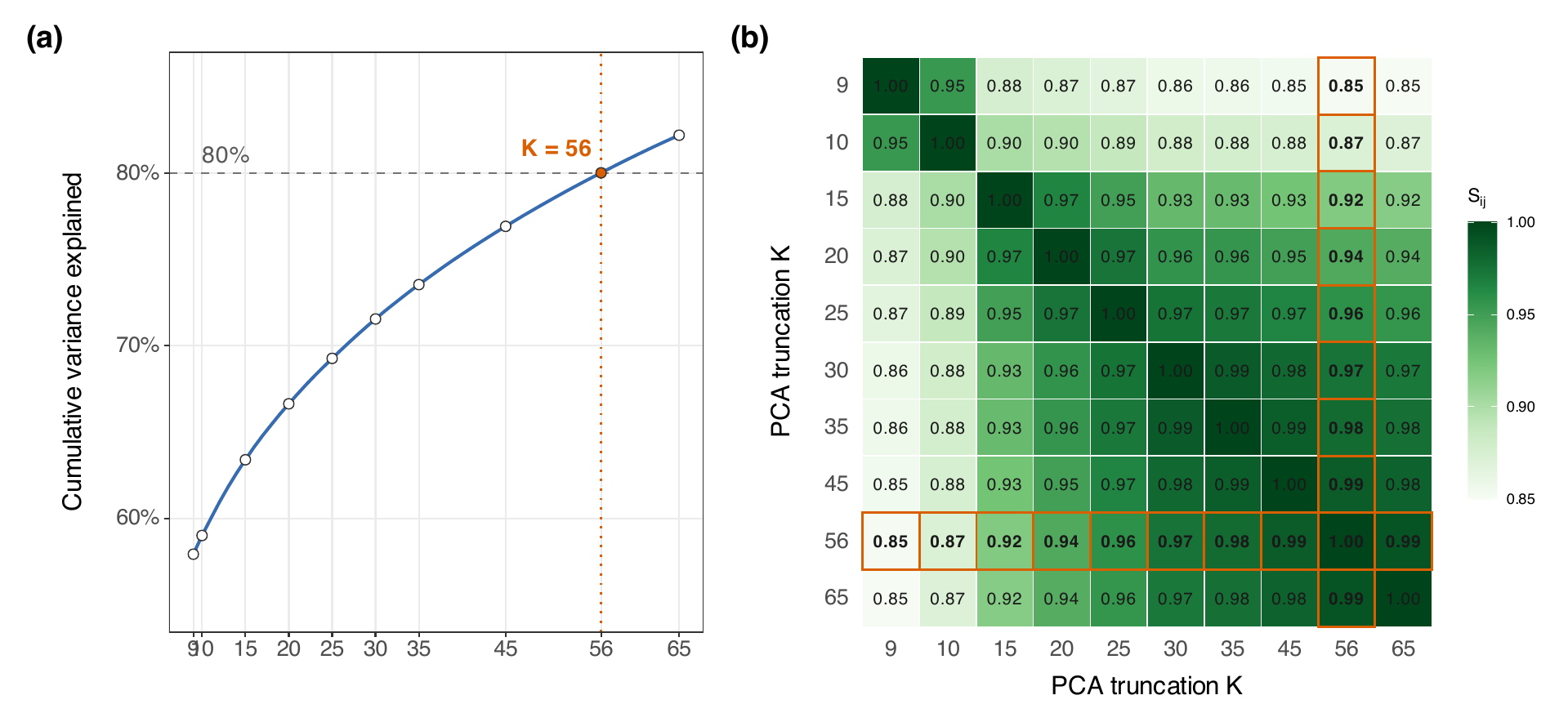}
		\caption{
			\textbf{Sensitivity of the multiCAsp solution to
				gene-expression PCA truncation.}
			Panel (a) shows the cumulative proportion of gene-expression variance explained as a function of the retained PCA dimension $K$, with the 80\%-variance threshold corresponding to $K=56$.
			Panel (b) shows the pairwise similarity between multiCAsp score subspaces
			estimated at different PCA truncation dimensions. The row and column corresponding to the
			80\%-variance choice ($K=56$) are outlined in orange.
		}
		\label{fig:ge_k_stability}
	\end{figure}
	
	\begin{figure}[!htbp]
		\centering
		\includegraphics[
		width=\linewidth,
		height=0.45\textheight,
		keepaspectratio
		]{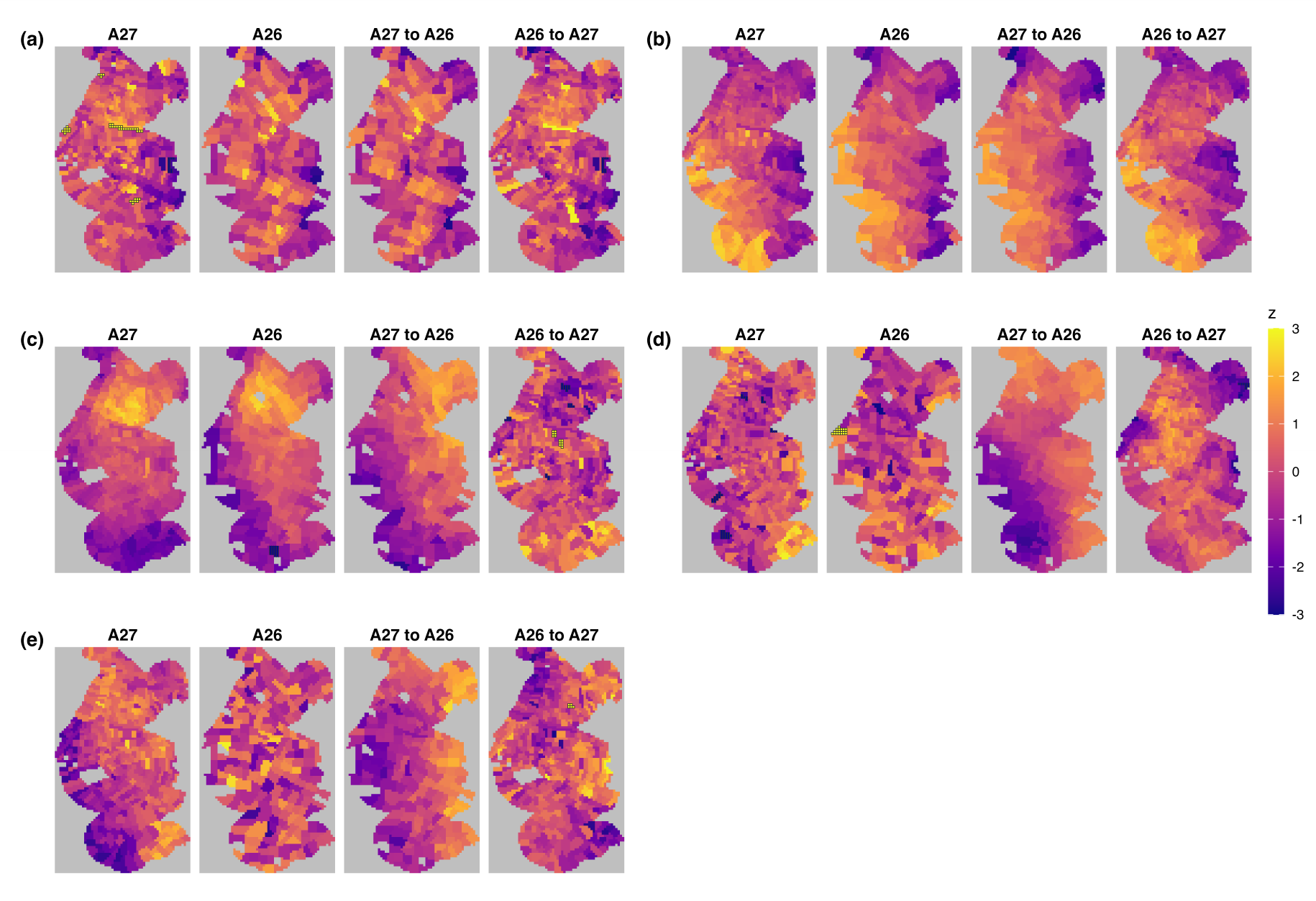}
		\caption{
			\textbf{Bidirectional cross-animal transfer of shared spatial factors using metabolomic measurements alone.}
			Panels (a)--(e) correspond to Factors 1--5. Within each panel, the four columns show independently estimated factors in A27 and A26, followed by transferred factors from A27 to A26 and from A26 to A27, respectively. Spatial factors were transferred using fixed source metabolomic loadings and target metabolomic measurements. }
		\label{fig:multicasp_bidirectional_transfer}
	\end{figure}
	
	\begin{figure}[!htbp]
		\centering
		\includegraphics[
		width=\linewidth,
		height=0.34\textheight,
		keepaspectratio
		]{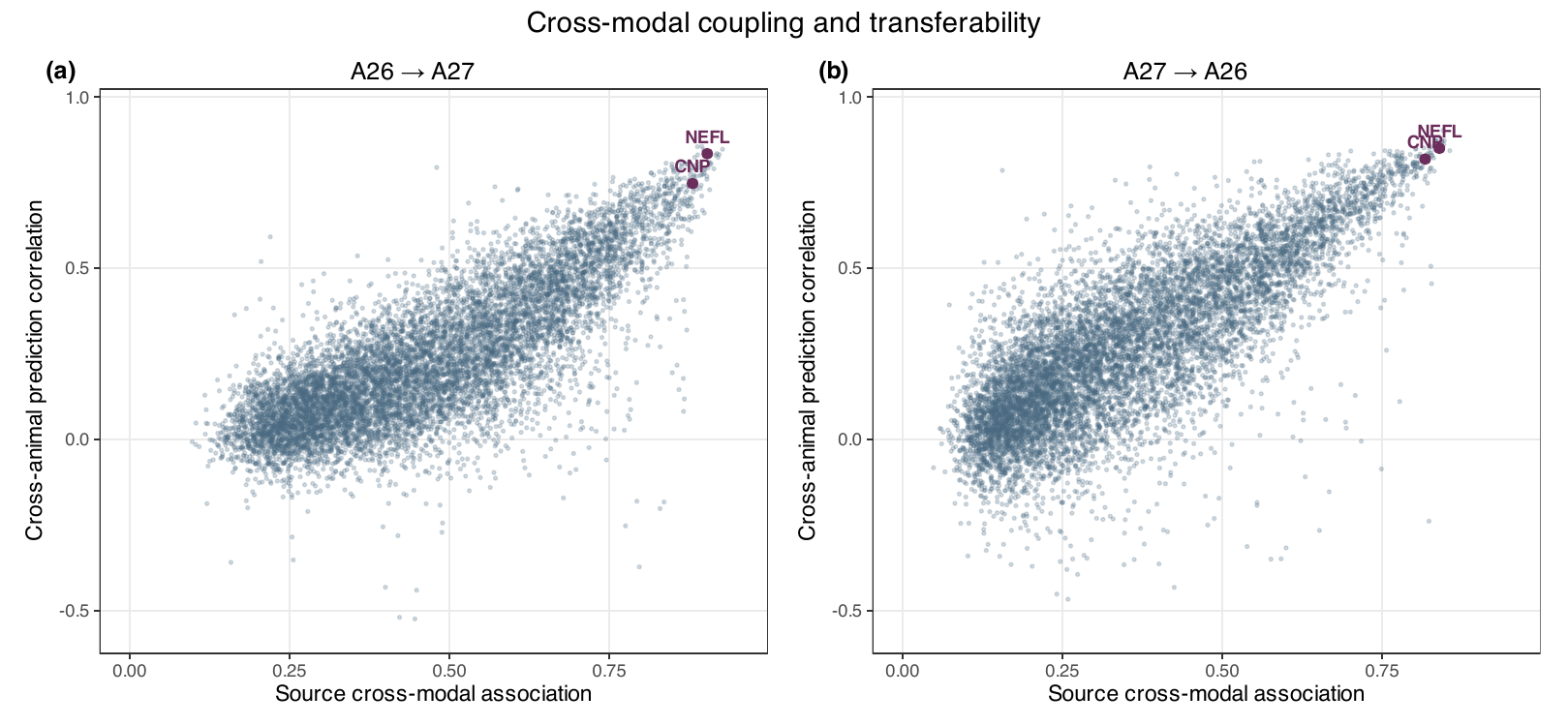}
		\caption{\textbf{Source cross-modal coupling and cross-animal protein prediction.} Each point represents a protein shared between A26 and A27. Panels (a) and (b) show transfer from A26 to A27 and from A27 to A26, respectively. Source cross-modal association is plotted against
			cross-animal prediction correlation. NEFL and CNP are highlighted in both panels.}
		\label{fig:multicasp_coupling}
	\end{figure}
	
	\clearpage
\end{document}